\documentclass[aps, pra, twocolumn, superscriptaddress, longbibliography,floatfix]{revtex4-2}
\usepackage{graphicx}
\usepackage{amsmath}
\usepackage{amssymb}
\usepackage{amsfonts}
\usepackage{color,xcolor}

\begin{document}

\title{Terahertz measurements on subwavelength-size samples down to the tunneling limit}

\author{D. Maluski}
\affiliation{II. Physikalisches Institut, Universit\"{a}t zu K\"{o}ln, D-50937 K\"{o}ln, Germany}
\author{I. C\'amara Mayorga}
\affiliation{Max-Planck-Institut für Radioastronomie, Auf dem H\"{u}gel 69, 53121 Bonn, Germany}
\author{J. Hemberger}
\author{M. Gr\"{u}ninger}
\affiliation{II. Physikalisches Institut, Universit\"{a}t zu K\"{o}ln, D-50937 K\"{o}ln, Germany}

\date{March 22, 2022}

\begin{abstract}
For terahertz spectroscopy on single crystals, the wavelength $\lambda$ often is comparable to the size of the studied samples, emphasizing diffraction effects. 
Using a continuous-wave terahertz spectrometer in transmission geometry, 
we address the effect of the sample size on the achievable accuracy of the optical properties, 
focusing in particular on the phase data. 
We employ $\alpha$-lactose monohydrate as a paradigmatic example and compare data that were measured using apertures with diameters $D$ in the range 
from 10\,mm to 0.2\,mm.
For small $D$, strong diffraction typically invalidates a quantitative analysis of the transmitted amplitude at low frequencies. The phase data, however, can be evaluated to lower frequency and show a more systematic dependence on $D$.
For a quantitative analysis, we employ a waveguide picture for the description of small apertures with 
a cylindrical bore.
For $D$ as small as 0.2\,mm, corresponding to $1/D$\,=\,50\,cm$^{-1}$, 
a circular waveguide does not support propagating waves below its cut-off frequency 
$1/\lambda_c$\,=\,$\omega_c/2\pi c \approx 29$\,cm$^{-1}$. 
Experimentally, we confirm this cut-off for cylindrical apertures with a thickness $d_{\rm ap}$\,=\,1\,mm.  Close to $\omega_c$, the measured phase velocity is an order of magnitude larger than $c$, the speed of light in vacuum. 
The cut-off is washed out if a sample is mounted on a thin aperture with a conical bore. 
In this case, the phase data of $\alpha$-lactose monohydrate for $D$\,=\,0.2\,mm can quantitatively be 
described down to about 10\,cm$^{-1}$ if the waveguide-like properties of the aperture are taken 
into account in the analysis.
\end{abstract}

\maketitle

\section{Introduction} 
\label{sec:introduction}

Terahertz spectroscopy keeps developing at an enormous pace in recent years \cite{Lee08,Tonouchi07,Jepsen11,Baxter11,Hughes12,OHara12,Zouaghi13,Lewis14,Safian19}. Measurements up to several THz are easily achieved today and usage of the term 'terahertz gap' smells old-fashioned. For broadband investigations of solid-state samples, however, the low-frequency regime -- building a bridge to the microwave range -- still is highly demanding. Above about 20\,GHz, standard microwave techniques become increasingly difficult because of 
semiconductor electronics suffering from calibration issues related to the very small waveguide geometries 
\cite{Scheffler05,Krupka06}. Terahertz setups typically address the range above about 100\,GHz \cite{Lee08,Tonouchi07,Jepsen11,Baxter11,Hughes12,OHara12,Zouaghi13,Lewis14,Safian19} equivalent to a vacuum wavelength $\lambda$ of about 3\,mm but the available size of, e.g., single crystalline samples is often much smaller. 

Terahertz measurements below the Rayleigh diffraction limit can be realized using near-field methods such as scanning the sample with a metallic tip while illuminating it with terahertz radiation \cite{Kersting05,Ribbeck08,Stinson18},
which is demanding both experimentally and in terms of the sophisticated data analysis. Further tools for sub-wave\-length terahertz imaging include laser THz emission microscopy \cite{Inoue06}, the creation of a plasma channel in air using a femtosecond laser pulse \cite{DAmico07,Zhao14}, or wavefront manipulation of a photoconductive terahertz emitter \cite{Baillergeau16}. Sub-wavelength focusing can also be achieved by left-handed metamaterials \cite{Pendry00,Grbic04}, which typically operate in a narrow frequency range. Using a 'standard' broadband continuous-wave (cw) setup, we demonstrate experimentally that the phase data can be evaluated to lower frequencies than the transmittance and that the phase offers a reasonable estimate of the optical properties for wavelengths much larger than the sample size if the properties of the employed aperture are taken into account in the analysis. Strong diffraction for sample sizes comparable to the wavelength severely affects the electric field amplitude measured at the detector, but those photons that still arrive at the detector carry reasonable phase information. As an example, we report on data of $\alpha$-lactose monohydrate measured on apertures with a diameter $D$ down to 0.2\,mm.

In passing, we study the terahertz properties of the circular apertures that are used as sample holders for the measurements on lactose. We aim in particular at a quantitative understanding of the phase data. The transmission through sub-wavelength apertures was already addressed by Bethe \cite{Bethe44}, who e.g.\ found that the transmission is low because of the poor coupling to radiative electromagnetic modes. 
Such sub-wavelength apertures attracted enormous interest when extraordinary optical transmission 
was discovered in periodic hole arrays in thin metallic films \cite{Ebbesen98,GarciaVidal2010}. 
At very low frequencies, a single aperture with a cylindrical bore acts like an undersized waveguide \cite{Gallot2000}.  
Below a cut-off frequency $\omega_c$, such a waveguide supports evanescent modes rather than propagating waves. Accordingly, light has to tunnel through the aperture. 
Above $\omega_c$, the propagating modes show a strong dispersion following 	
$\omega(q) = \sqrt{\omega_c^2 + (q\,c)^2 }$, 
where $q$ denotes the momentum and $c$ the speed of light. For $\omega$ only slightly larger than $\omega_c$, this dispersion gives rise to a pronounced enhancement of the phase velocity to values much larger than $c$.
This behavior is well-known for waveguides in the microwave range \cite{Collin} where it has been discussed controversially as a signature of superluminal effects \cite{Nimtz92,Winful06,With10,Nimtz14}.
In the terahertz range, a similar waveguide approach has been used to detect absorption signatures in microliter DNA solutions \cite{Zhang13}, where the liquid sample has been placed \textit{inside} the waveguide, in contrast to our setup.
We focus on the phase data of small cylindrical apertures and demonstrate that these are well described by the waveguide picture. 
Moreover, we show that the phase data of $\alpha$-lactose monohydrate measured on a small aperture with $D$\,=\,0.2\,mm, where the cylindrical bore partially has been widened to a cone, systematically deviates from 
the result obtained on a large aperture but that this behavior can be described reasonably well by taking into account the waveguide character of the cylindrical part of the aperture.

In the following, we first address the terahertz properties of $\alpha$-lactose monohydrate as determined using a large aperture. Then we study the properties of bare apertures of different diameters $D$, comparing apertures with a cylindrical bore with partially conical ones.
Finally, we employ the latter to measure $\alpha$-lactose monohydrate for $D$ down to 0.2\,mm.

\section{Setup} 
\label{sec:setup}

\begin{figure}[t]
	\centering
	\includegraphics[width=1.0\columnwidth]{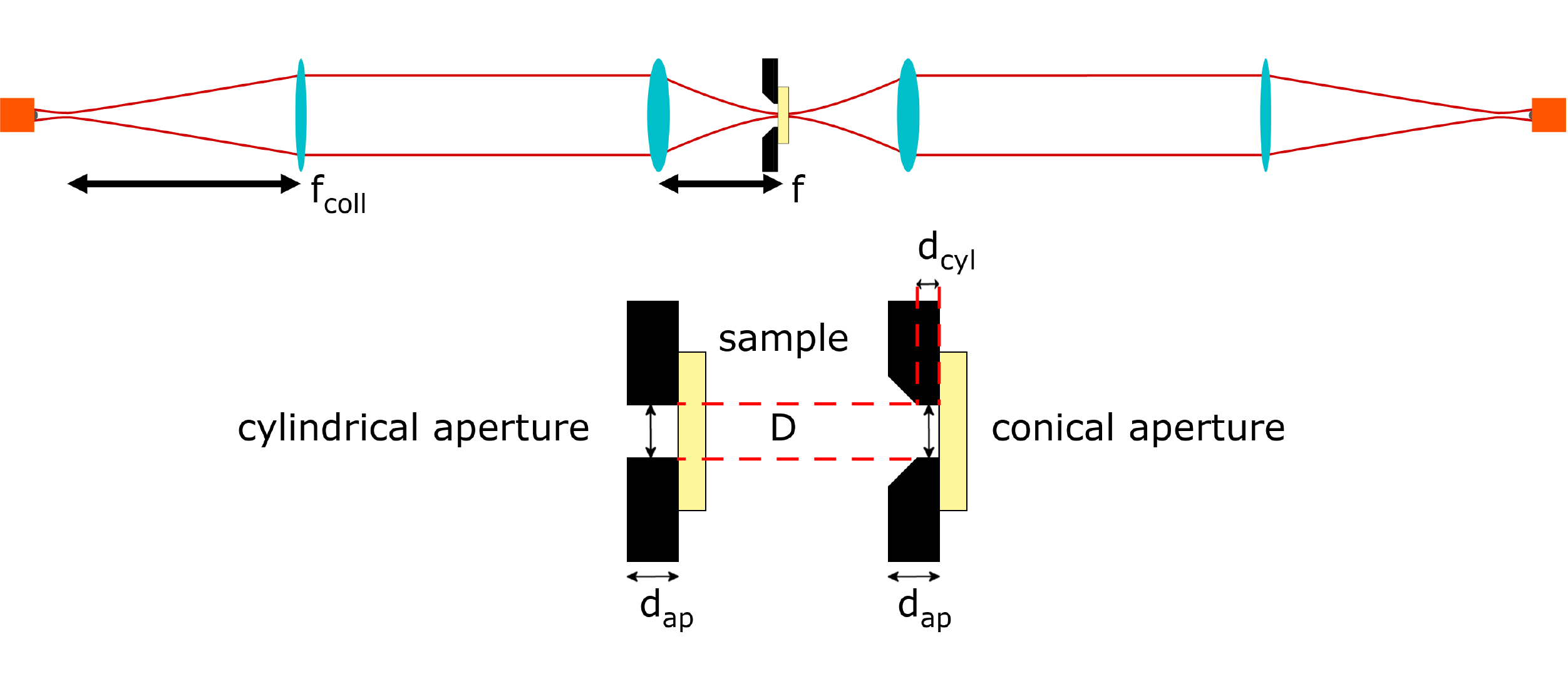}
	\caption{Top: Four-lens setup equivalent to two Gaussian telescopes with $f_{\rm coll}$\,=\,20\,cm and $f$\,=\,10\,cm. 
		The lens radius equals 5\,cm. Orange squares denote THz source (left) and phase-sensitive detector, respectively. 
		The aperture for mounting the sample is shown in black. Bottom: Sketch of the two aperture types. 
	The cylindrical apertures have a cylindrical bore with diameter $D$, and their thickness equals $d_{\rm ap}$. For the conical apertures, the cylindrical bore 
	is partially widened to a conical shape with an opening angle of $90^\circ$. Then, $d_{\rm ap}$ refers to the total thickness whereas $d_{\rm cyl}$ refers to the thickness of the remaining cylindrical part. The sample is a pressed pellet of $\alpha$-lactose monohydrate with plane-parallel surfaces.
	}
	\label{fig:setup}
\end{figure}

We employ cw terahertz spectroscopy based on photomixing with phase-sensitive homodyne detection. 
Using the beat of two tunable near-infrared lasers ({\sc Toptica}), the system covers 
the frequency range of about 3\,cm$^{-1}$ to 50\,cm$^{-1}$ with a frequency resolution lying in the MHz range. 
The photomixers ({\sc Max-Planck-Institut für Radioastronomie}) employ log-spiral antennae which emit and 
detect circularly polarized light over a broad frequency range.
The setup is equipped with two fiber stretchers that act as fast phase modulators, 
allowing us to determine both amplitude and phase of the terahertz radiation with high accuracy 
and a large dynamic range. 
Details of the setup were described in Refs.\ \cite{Roggenbuck10,Roggenbuck12,Roggenbuck13}, 
and a quantitative description of the overall phase behavior in our setup was reported in Ref.\ \cite{Langenbach14}, 
focusing on the photomixers and the log-spiral antennae. 
If necessary, the accuracy of the phase data can be further enhanced by performing self-normalizing phase measurements based on photomixing of three lasers, which eliminates (thermal) drifts of the setup \cite{Komu15}. All data presented here were collected using the standard two-laser setup.

\begin{figure}[t]
	\centering
	\includegraphics[width=0.95\columnwidth]{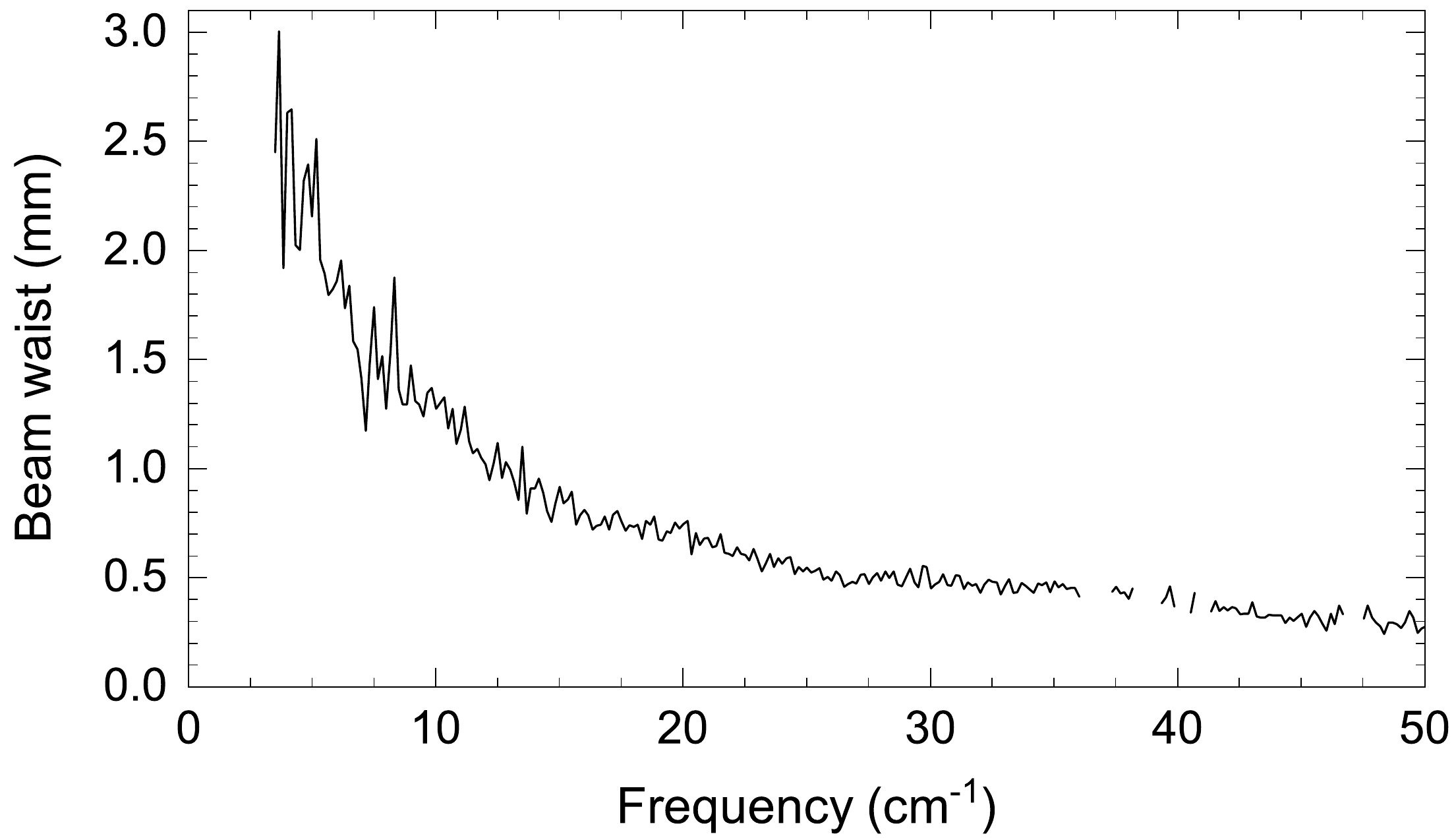}
	\caption{Beam waist $w_0$\,=\,$w_e/2$ at the sample focus position, measured with the knife-edge technique. 
		The beam was cut off by moving a razor blade perpendicular to the propagation direction. 
		The analysis assumes a Gaussian beam profile.}
	\label{fig:beamwaist}
\end{figure}

The photomixers are equipped with hyperhemispherical Si lenses. 
The optics uses four hyperbolic biconvex lenses with a radius of 5\,cm to focus the emitter waist  first on the sample and then on the receiver waist, see Fig.\ \ref{fig:setup}. 
The two outer, collimating lenses facing source and detector have a focal length $f_{\rm coll}$\,=\,20\,cm while the two inner lenses focusing on the sample show $f$\,=\,10\,cm. 
A Gaussian telescope configuration was employed, i.e., the distance between inner and outer lenses 
amounts to $f_{\rm coll}+f$. Thereby the magnification and output waist position are independent of the wavelength. 
The input Gaussian telescope scales down the emitter waist by a factor $f/f_{\rm coll}$\,=\,1/2 at the sample position. 
A small waist thus is obtained for large $f_{\rm coll}$ and small $f$. 
According to electromagnetic simulations of the beam characteristics of our photomixers, we expect a beam waist $w_e$ of the emitter of ($2.8\pm 0.4$)\,mm at 10\,cm$^{-1}$, ($2.4\pm 0.4$)\,mm at 15\,cm$^{-1}$, and 
($1.5\pm 0.2$)\,mm at 33\,cm$^{-1}$. 
These values roughly agree with measurements of the beam waist $w_0$\,=\,$w_e/2$ at the sample position  
as measured with the knife-edge technique, see Fig.\ \ref{fig:beamwaist}. 
The beam waist $w_0$ decreases with increasing frequency and drops below 1\,mm (0.5\,mm) at about 13\,cm$^{-1}$ (26\,cm$^{-1}$).
The output Gaussian telescope transforms the waist back to its original dimensions for optimal beam 
matching at the receiver. The emitter/receiver waist positions were placed at the focal points of the 
input/output (collimating) lenses.

To study the effect of the aperture size, we employ commercially available powder of $\alpha$-lactose monohydrate with well-known terahertz properties \cite{Roggenbuck10}. Using a pellet press with a pressure of 6\,kbar, the powder was pressed into a pellet with a diameter of 13\,mm, a thickness of 0.85\,mm, and plane-parallel surfaces. For mounting the samples, we employ Cu apertures with a thickness $d_{\rm ap}$ in the range from 0.5 to 1.5\,mm. Here, we compare apertures with a cylindrical bore with ones where the initial cylindrical bore is partially widened to a conical shape 
on one face, see Fig.\ \ref{fig:setup}.
Usually, we use such partially conical bores, which for the sake of simplicity we call conical apertures in the following.
The sample is glued on top of the cylindrical side of the bore, 
see Fig.\ \ref{fig:setup}. This is the typical configuration for transmittance measurements of (large) single crystals. In the free-beam limit, where the beam waist $w_0$ is much smaller than the aperture diameter, the aperture has no effect on the measurement. Our aim is to study and understand the deviations from the free-beam limit at low frequencies for small samples.

\section{Data of large samples}
\label{sec:datalargesamples}

\begin{figure}[t]
	\centering
	\includegraphics[width=0.88\columnwidth]{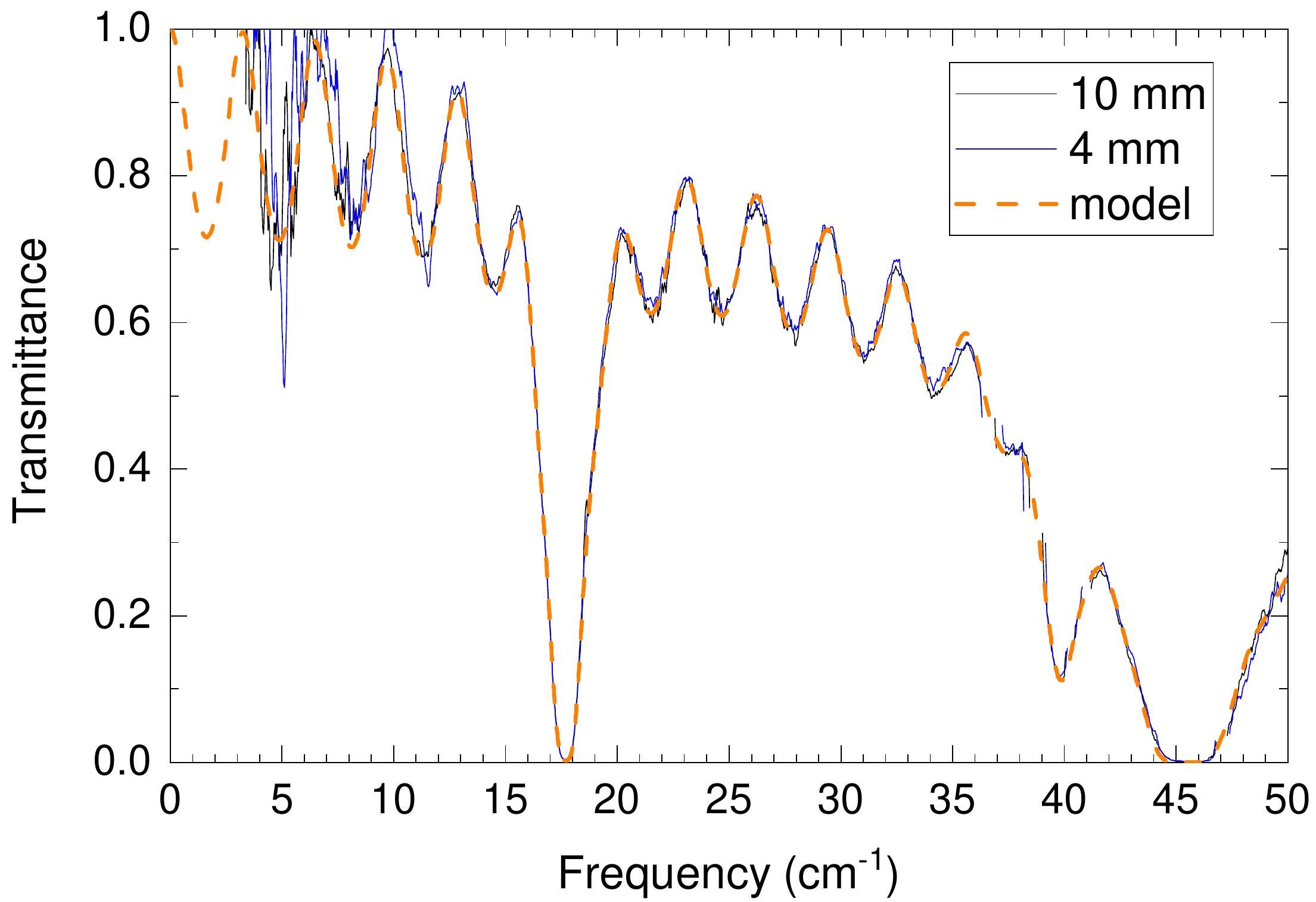}
	\hspace*{3mm}
	\includegraphics[width=1\columnwidth]{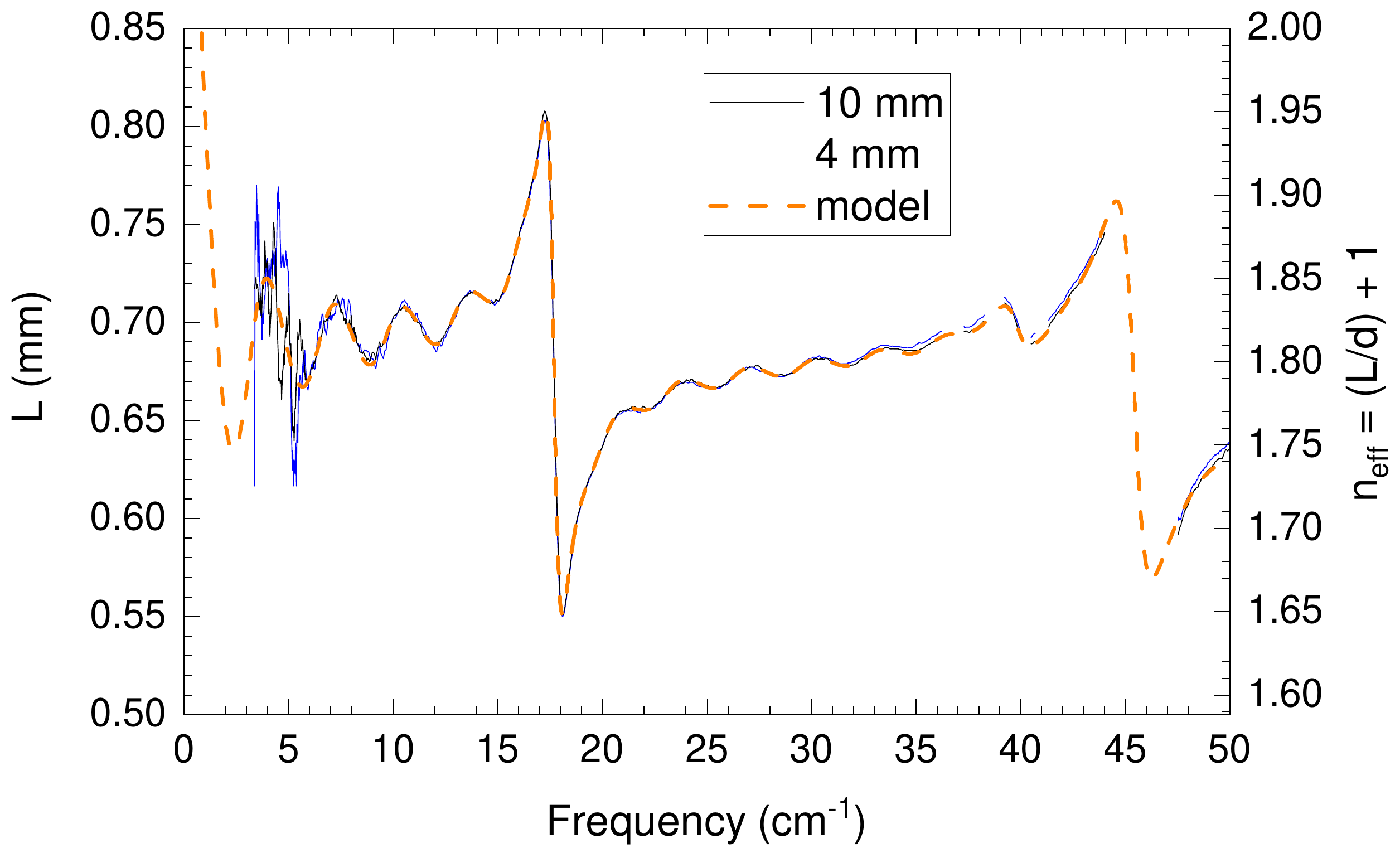}
	\caption{Transmittance $T(\omega)$ (top) and terahertz path length difference 
		$L(\omega)$\,=\,$(\varphi/2\pi)\cdot \lambda$ (bottom left axis) of $\alpha$-lactose monohydrate measured with aperture diameters of $D$\,=\,10\,mm and 4\,mm. 
		Bottom right axis: corresponding estimate of the refractive index using the single-bounce approximation $L$\,=\,$(n_{\rm eff}-1)\,d$.	
		Data points were collected with a step width of 1\,GHz\,$\approx$\,0.03\,cm$^{-1}$. 
		Small gaps in the measured frequency range, e.g.\ around 37\,cm$^{-1}$, are due to absorption bands of water vapor in the beam path \cite{Roggenbuck10}. 
		Additionally, the phase (and hence $L(\omega)$) cannot be determined when the amplitude drops below the noise level, for instance around the absorption 
		band at 45\,cm$^{-1}$.
		Orange curve in both panels: Drude-Lorentz fit, which yields a sample thickness  $d$\,=\,0.85\,mm.  
	}
	\label{fig:Transmission_DeltaL_fit}
\end{figure}

As starting point, we address the transmittance $T(\omega)$ of a pressed pellet of 
$\alpha$-lactose monohydrate measured under ideal conditions, i.e., the sample is glued 
onto an aperture with diameter $D$ much larger than the beam waist $w_0$. 
In this case, the details of the aperture are irrelevant. The two data sets for $D$\,=\,10\,mm and 4\,mm agree very well with each other over the full measured frequency range, 
see top panel of Fig.\ \ref{fig:Transmission_DeltaL_fit}. 
We observe the three well-known absorption features \cite{Roggenbuck10} at about 17.7\,cm$^{-1}$, 39.8\,cm$^{-1}$, and 45.4\,cm$^{-1}$. In the frequency ranges showing high transmittance, the data exhibit pronounced Fabry-Perot interference fringes caused by multiple reflections within the sample.

The bottom panel of Fig.\ \ref{fig:Transmission_DeltaL_fit} depicts the measured data of the phase $\varphi(\omega)$ in terms of the terahertz path length difference $L$\,=\,$(\varphi/2\pi)\,\cdot \lambda$ induced by the sample as compared to air, i.e., an empty aperture used for the reference measurement. 
Akin to the transmittance, the data of $L(\omega)$ for $D$\,=\,10\,mm and 4\,mm agree very well with each other. 
In first approximation, considering only the ray passing the sample once while neglecting internal reflections within the 
sample, the terahertz path length difference is a measure of the real part 
$n(\omega)$ of the refractive index. 
With the sample thickness $d$, the effective quantity $n_{\rm eff}(\omega)$\,=\,$1+(L/d)$  
offers a reasonable first estimate of the correct $n(\omega)$. 
The main difference arises due to the presence of interference fringes in 
$L(\omega)$ and $n_{\rm eff}(\omega)$, 
see right axis of the bottom panel of Fig.\ \ref{fig:Transmission_DeltaL_fit}.

A precise result of the \textit{complex} refractive index  $N(\omega)$ 
=\,$n(\omega)+{\rm i}\,k(\omega)$, or, equivalently, the complex dielectric function $\varepsilon(\omega)$
can be obtained via a fit using a  Drude-Lorentz oscillator model. 
A simultaneous fit of $T(\omega)$ and $L(\omega)$ also yields a reasonable value of the sample thickness $d$. This is possible since the extrema of the interference fringes depend on $n\cdot d$, whereas $L(\omega)$ is a measure of $(n-1)\cdot d$. 
The corresponding fit result describes the experimental data accurately, see dashed orange lines in both panels of Fig.\ \ref{fig:Transmission_DeltaL_fit}. 
The fit yields a sample thickness $d$\,=\,0.85\,mm, which agrees within about 2\% with measurements in an optical microscope.

For known sample thickness $d$, similar results can be obtained by separate fits of either the transmittance or the phase data, i.e., $L(\omega)$.  
Analyzing the phase data, a small error is caused by neglecting the Gouy phase shift \cite{Kuzel10,Liang15}. 
Compared to a plane wave, a Gaussian beam that passes through a beam waist experiences a phase shift of $\pi$. 
Even though this applies to the measurements of both, sample and reference, 
it should be considered in the data analysis, see Appendix. 
However, a quantitative estimate for our case shows that the corresponding error of $n_{\rm eff}$ 
is less than 0.05 (0.01) above 3\,cm$^{-1}$ (10\,cm$^{-1}$), as discussed in the Appendix.  
This small error is negligible for the present discussion of small apertures.

\section{Comparing different apertures}
\label{sec:differentapertures}

\subsection{Cylindrical bore}
\label{sec:cylindricalbore}

\begin{figure}[t!]
	\centering
	\includegraphics[width=\columnwidth]{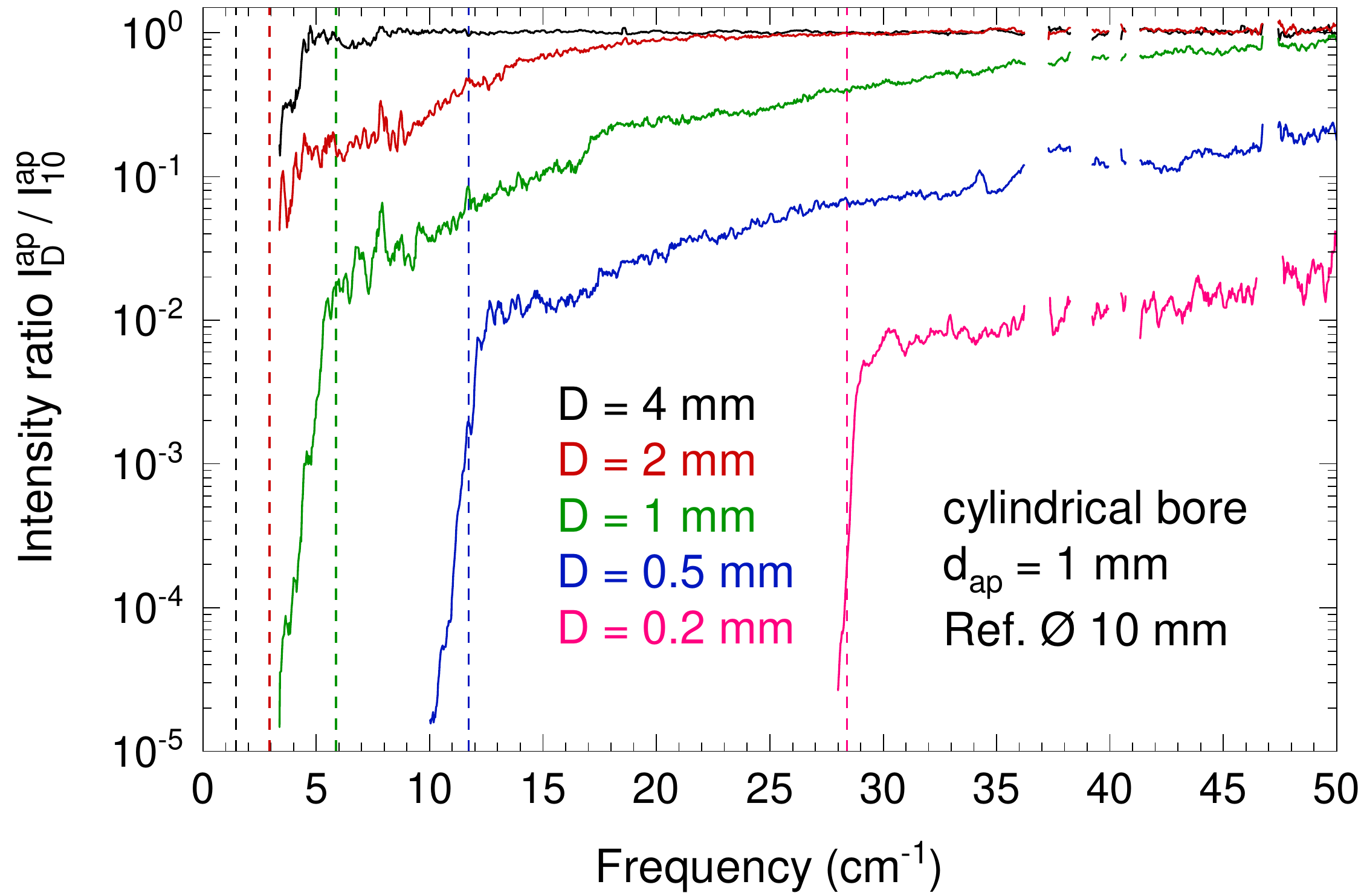}
	\includegraphics[width=\columnwidth]{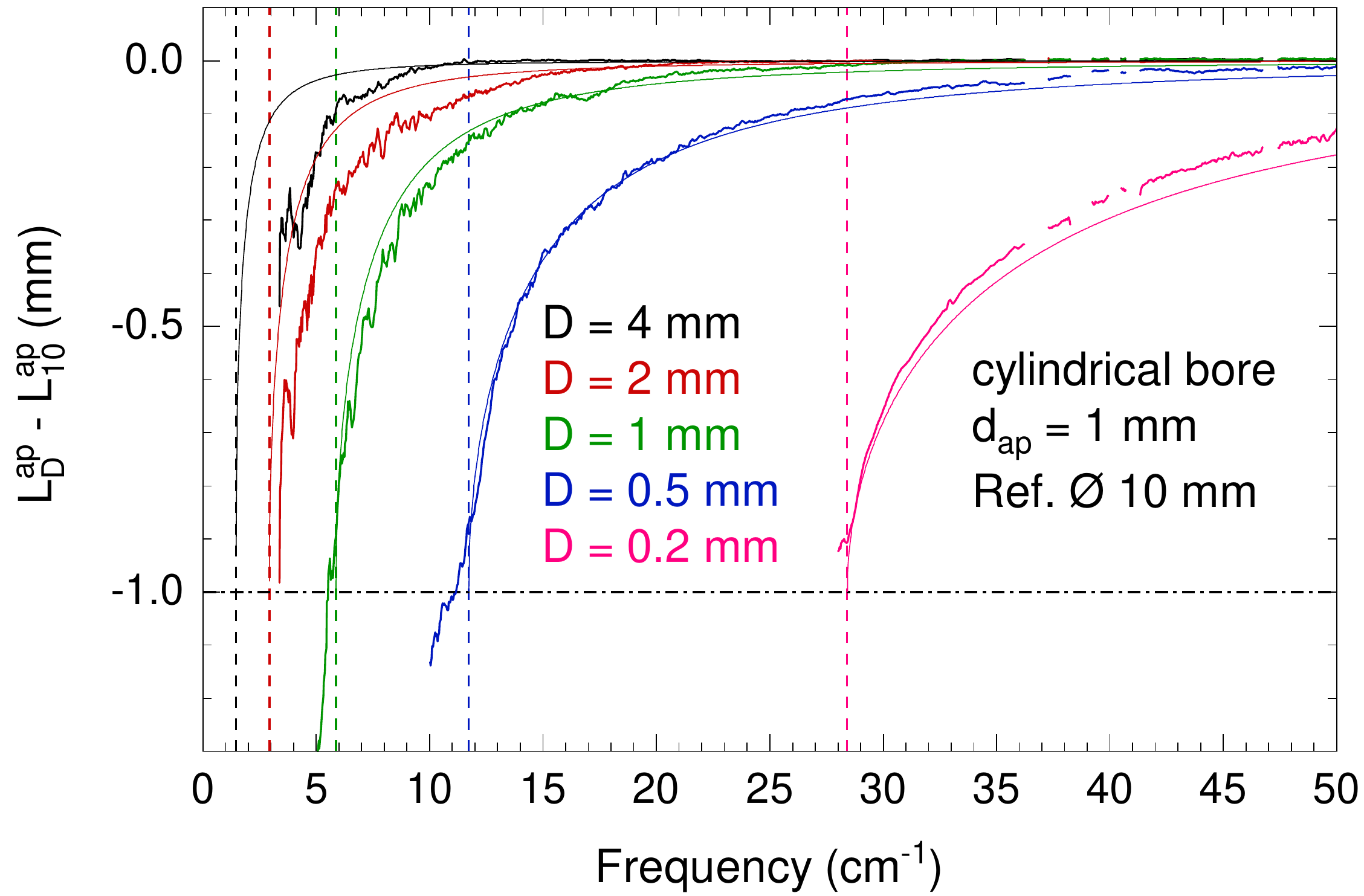}
	\caption{Measured intensity ratio $I^{\rm ap}_D/I^{\rm ap}_{10}$ (top) and terahertz path length difference $L^{\rm ap}_D-L^{\rm ap}_{10}$ (bottom) for different apertures with cylindrical bore with diameter $D$ and thickness $d_{\rm ap}$\,=\,1\,mm, normalized to a reference with $D$\,=\,10\,mm. 
	In both panels, vertical dashed lines denote the waveguide cut-off frequency $\omega_c$ for the curves of the same color. Below $\omega_c$, the apertures do not support propagating waves. For the smallest aperture with nominally $D$\,=\,0.2\,mm, $\omega_c$ was reduced by 3\,\% to correct for the actual size of $D$. 	
	In the bottom panel, thin solid lines depict the prediction of Eq.\ (\ref{eq:DL}), and the horizontal dash-dotted line denotes $-d_{\rm ap}$. 
	Above 35\,cm$^{-1}$, the narrow features due to water vapor were cut out from the data. 
}
\label{fig:refs_Trans_DeltaL}
\end{figure}

Before addressing the measurements of a sample on a small aperture, we first compare the empty apertures directly, 
using the data for $D$\,=\,10\,mm as a reference for the smaller apertures. 
In the following, $I^{\rm ap}_D$ and $L^{\rm ap}_D$ refer to data of the intensity and the terahertz path length difference, respectively, measured on empty apertures.
For a cylindrical bore, Fig.\ \ref{fig:refs_Trans_DeltaL} depicts the corresponding intensity ratios $I^{\rm ap}_D/I^{\rm ap}_{10}$ as well as $L^{\rm ap}_D(\omega)-L^{\rm ap}_{10}(\omega)$, i.e., the difference of the terahertz path length difference $L^{\rm ap}_D$ for an aperture with diameter $D$ and the one of the 10\,mm aperture. 
This figure reveals clear signatures of three different frequency ranges: 
We observe free-beam optics at high frequencies where the beam waist $w_0$ is smaller than roughly $D/3$.
At low frequencies, we find a steep suppression of the intensity $I^{\rm ap}_D$ at a cut-off frequency $\omega_c \propto 1/D$ (dashed lines) below which the aperture does not support propagating waves. 
For instance for $D$\,=\,1\,mm, we roughly find $I^{\rm ap}_D/I^{\rm ap}_{10}$\,$\propto \omega^{12}$ below 6\,cm$^{-1}$. 
In the intermediate range between $\omega_c$ and the high-frequency limit, $I^{\rm ap}_D/I^{\rm ap}_{10}$ 
is gradually suppressed with decreasing frequency while the phase data for small $D$ can quantitatively be described in terms of modes propagating in a waveguide. In the following, we take a closer look at these three ranges.

\begin{figure}[t]
	\centering
	\includegraphics[width=0.93\columnwidth]{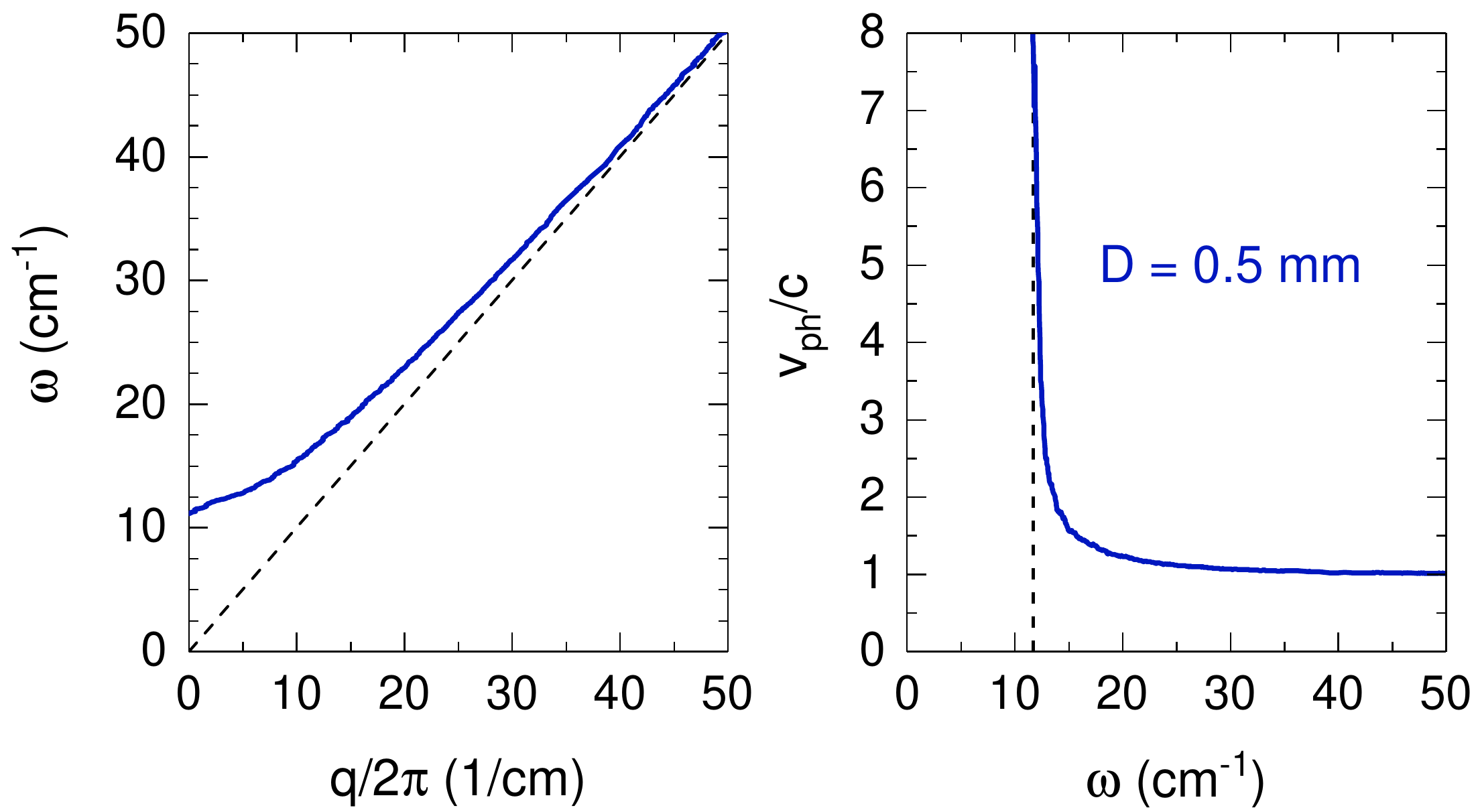}
	\caption{Dispersion $\omega(q)$ (left) and phase velocity $v_{\rm ph}$ (right) for waves propagating in the 
		cylindrical aperture with $D$\,=\,0.5\,mm, as determined from the measured terahertz path length difference $L^{\rm ap}_{0.5}$ and Eq.\ (\ref{eq:DL}). The dashed line in the left panel depicts the dispersion in free space, while the dashed line in the right panel shows the cut-off frequency $\omega_c$. 
	}
	\label{fig:dispersion}
\end{figure}

In the measured frequency range, the free-beam high-frequency limit is only reached for the large apertures with $D$\,=\,2\,mm and 4\,mm.  For these two, we find $I^{\rm ap}_D/I^{\rm ap}_{10}$ $\approx 1$ and 
$L^{\rm ap}_D(\omega)-L^{\rm ap}_{10}(\omega)$\,$\approx 0$ at high frequencies. As expected, apertures much larger than the beam waist have a negligible effect on both the transmitted intensity and the terahertz path length.

Driven by the frequency dependence of the beam waist, see Fig.\ \ref{fig:beamwaist}, 
the intensity ratio $I^{\rm ap}_D/I^{\rm ap}_{10}$ is gradually suppressed for frequencies below roughly 10\,cm$^{-1}$ for $D$\,=\,4\,mm and below about 20\,cm$^{-1}$ for $D$\,=\,2\,mm. A beam waist $w_0$ comparable to $D/3$ 
results in partial blocking of the incident beam as well as enhanced diffraction.
In this frequency range below the free-beam limit, $L^{\rm ap}_D(\omega)-L^{\rm ap}_{10}(\omega)$ is \textit{negative}, which means that an aperture with small $D$ causes a \textit{decrease} of the terahertz path length as compared to vacuum (or actually air). 
This is equivalent to a phase velocity $v_{\rm ph}$ larger than $c$, the speed of light.  	
For media, the terahertz path length difference $L(\omega)$ is a measure of the refractive index, as discussed above. 
In a frequency range where a given medium is transparent, the real part $n$ is larger than the vacuum value 1, such that the medium enhances the terahertz path length compared to vacuum. The stunningly different, opposite behavior observed in the bottom panel of Fig.\ \ref{fig:refs_Trans_DeltaL} is well known for (undersized) wave\-guides \cite{Gallot2000,Collin,Nimtz92}. 
Within a wave\-guide, the superposition of rays reflected 
from the walls yields a wavelength $\lambda_w$ which is \textit{larger} than the vacuum wavelength $\lambda$. 
Below a cut-off frequency $\omega_c = 2\pi c/\lambda_c \propto 1/D$ with a prefactor that is characteristic for each mode, the waveguide does not support propagating waves anymore and one expects evanescent modes. 
For a cylindrical waveguide, the TE$_{11}$ mode is the fundamental mode. It shows the lowest cut-off frequency $\omega_c \propto$ 
$1/\lambda_c$\,=\,1.841/$\pi D$, which yields, e.g., 11.7\,cm$^{-1}$ for $D$\,=\,0.5\,mm 
and 29.3\,cm$^{-1}$ for $D$\,=\,0.2\,mm (cf.\ vertical dashed lines in Fig.\ \ref{fig:refs_Trans_DeltaL}). 
The calculated cut-off frequencies coincide with the steep suppression of the intensity. 

In the vicinity of $\omega_c$ we find $L^{\rm ap}_D-L^{\rm ap}_{10}$\,$\approx -d_{\rm ap}$, i.e., the reduction of $L^{\rm ap}_D$ equals the thickness of the aperture, see Fig.\ \ref{fig:refs_Trans_DeltaL}.  
For evanescent modes, the real part of the wavevector {\bf q} vanishes, 
equivalent to a vanishing real part $n$ of the refractive index, infinite $\lambda_w$, and 
no spatial phase dependence. 
For an aperture of thickness $d_{\rm ap}$, 
the transmission of electromagnetic radiation via evanescent modes is equivalent to a tunneling process through the aperture. Evanescent modes do not propagate and hence 
the tunneling time vanishes and the apparent transmission time becomes independent of the geometrical length, 
the so-called Hartman effect \cite{Hartman62}. Over decades, this has given rise to fierce 
discussions on the possibility of superluminal effects for light pulses, 
see Refs.\ \cite{Winful06,Nimtz14} and references therein.
From our perspective of cw spectroscopy with a 'monochromatic' beam, the vanishing tunneling time 
below $\omega_c$ 
simply corresponds to a vanishing contribution of the aperture to the terahertz path length, equivalent to an infinite phase velocity within the bore. 
Compared to the case of a large aperture, this yields a reduction of $L^{\rm ap}_D$ by roughly $d_{\rm ap}$ at $\omega_c$ and below, in agreement with our experimental result.

\begin{figure}[t!]
	\centering
	\includegraphics[width=\columnwidth]{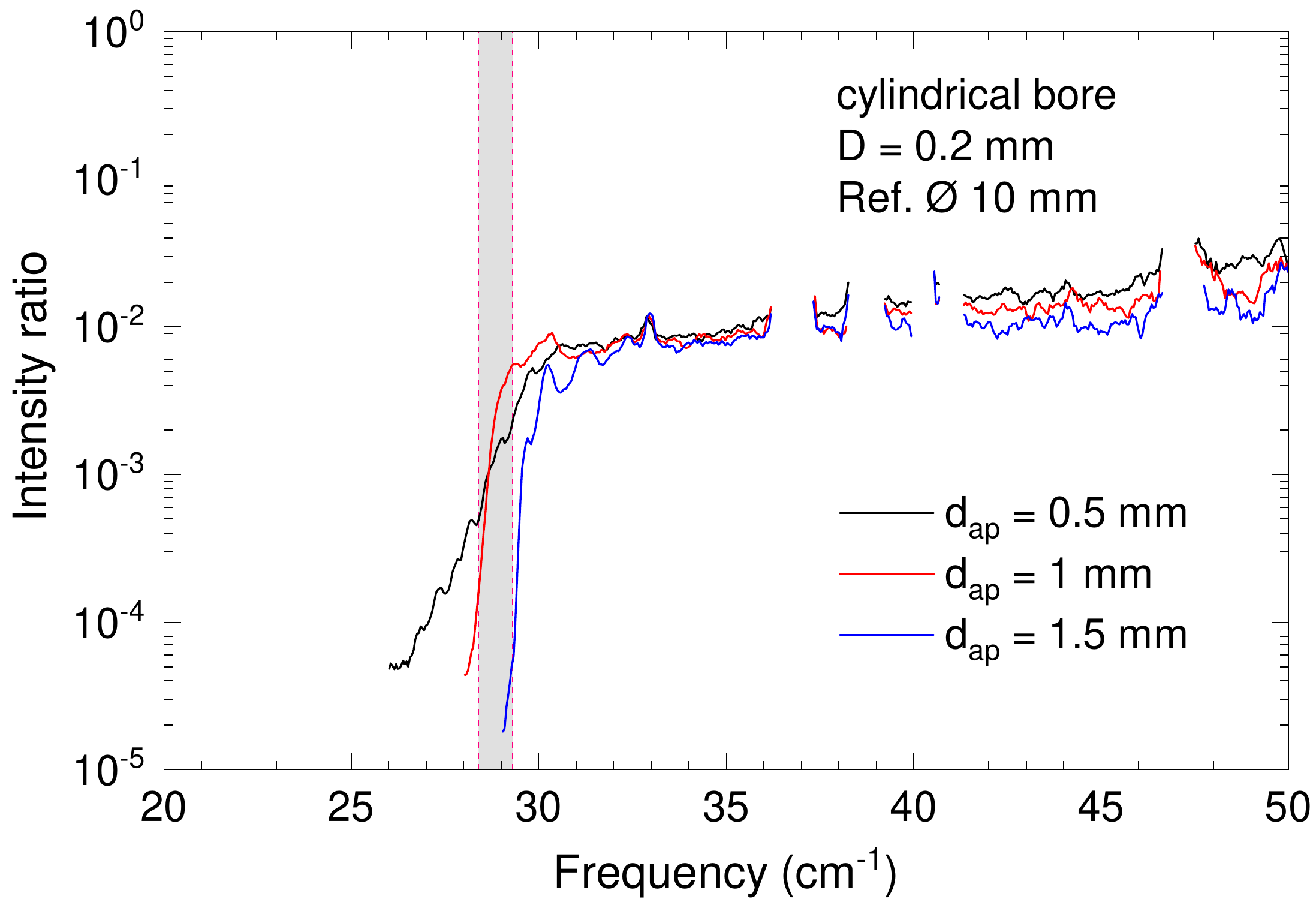}
	\includegraphics[width=\columnwidth]{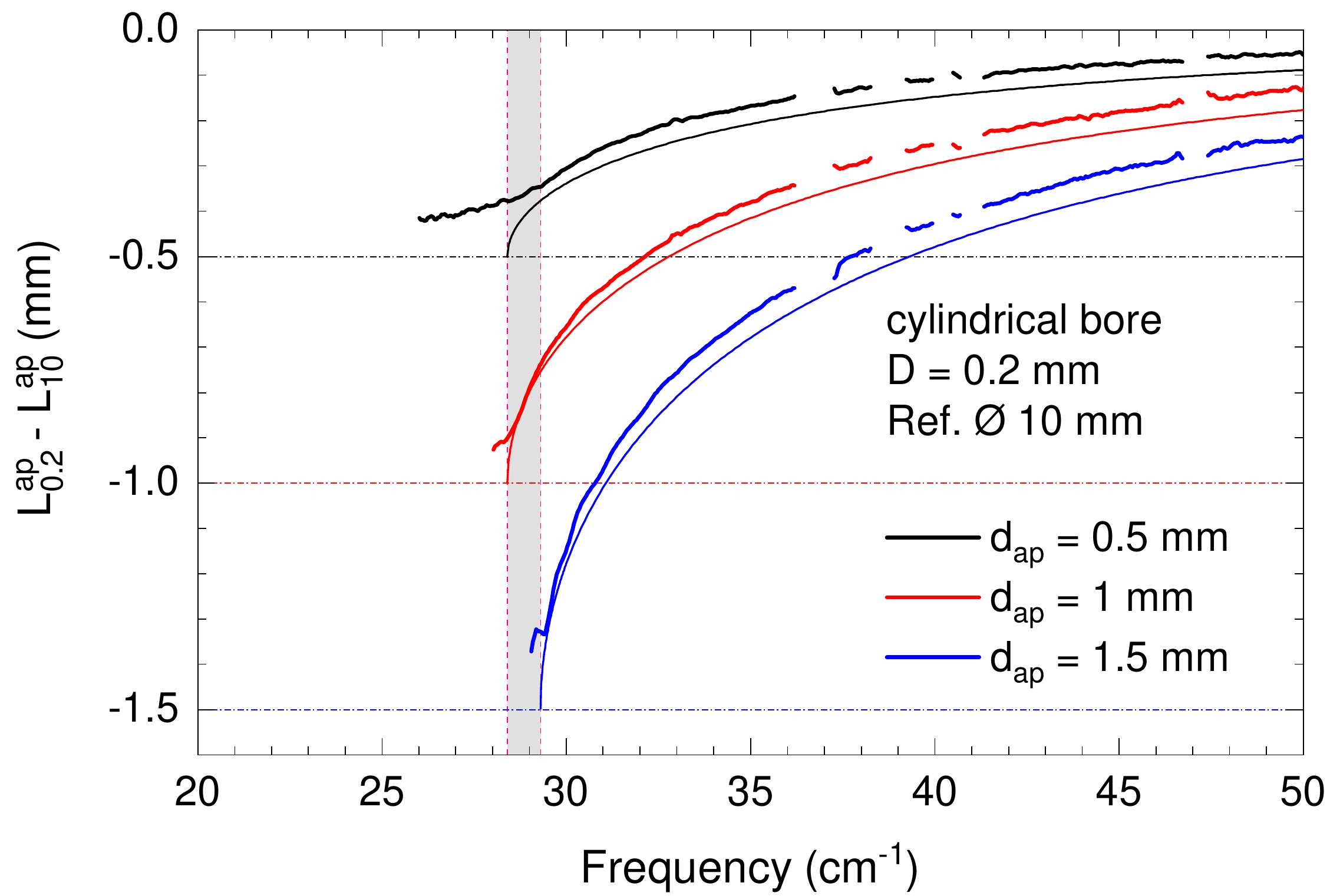}
	\caption{Intensity ratio $I^{\rm ap}_{0.2}/I^{\rm ap}_{10}$ (top) and terahertz path length difference $L^{\rm ap}_{0.2}-L^{\rm ap}_{10}$ (bottom) for different cylindrical apertures with nominal diameter $D$\,=\,0.2\,mm and different thicknesses $d_{\rm ap}$, normalized to a reference with $D$\,=\,10\,mm. 
	The cut-off frequency $\omega_c$ equals 29.3\,cm$^{-1}$ for $D$\,=\,0.2\,mm and is reduced to 
	28.4\,cm$^{-1}$ if $D$ is enhanced by 3\,\% (see gray area). 
	In the bottom panel, thin solid lines show predictions of Eq.\ (\ref{eq:DL}), and horizontal dash-dotted lines indicate $-d_{\rm ap}$. 
}
\label{fig:refs_02}
\end{figure}

Above $\omega_c$, a single mode in a waveguide shows strong dispersion according to 
$\omega(q) = \sqrt{\omega_c^2 + (q\,c)^2 }$ 
which yields a phase velocity $v_{\rm ph}$ given by 
\begin{equation}
\label{eq:vph}
\frac{v_{\rm ph}}{c} = \left[  1-\left(\frac{\omega_c}{\omega}\right)^2\, \right]^{-1/2}  \, ,
\end{equation}
a group velocity $v_{\rm g}$\,=\,$c^2/v_{\rm ph}$, and a terahertz path length difference
\begin{equation}
	\label{eq:DL}
	L^{\rm ap}_D = - d_{\rm ap}\! \left[  1 - \frac{c}{v_{\rm ph}}\right]\! =
	- d_{\rm ap}  \left[  1 - \sqrt{1-\left(\frac{\omega_c}{\omega}\right)^2} \right]  .
\end{equation}
The corresponding behavior of $L^{\rm ap}_D-L^{\rm ap}_{10}$ for the TE$_{11}$ mode is depicted by thin solid lines in the bottom panel of Fig.\ \ref{fig:refs_Trans_DeltaL} and describes our experimental data very well for $D \leq 1$\,mm\,=\,$d_{\rm ap}$. 
For $D$\,=\,2\,mm and 4\,mm, the aperture diameter $D$ is larger than the thickness $d_{\rm ap}$ and the quantitative applicability of the waveguide picture breaks down.

For $D\leq 1$\,mm, we may use the measured value of $L^{\rm ap}_D$ and Eq.\ (\ref{eq:DL}) to obtain an experimental result for $v_{\rm ph}$ as well as for $\omega(q)$. This is plotted in Fig.\ \ref{fig:dispersion} for the example of $D$\,=\,0.5\,mm, showing a phase velocity of about $8c$ in the vicinity of $\omega_c$.
A more detailed description of the data of empty apertures requires to take into account 
further waveguide modes, damping within the waveguide, as well as the coupling from and into free space, 
which is beyond the scope of the present work. 
Our aim was to qualitatively understand the different frequency ranges and to elucidate the behavior of the phase. 
We conclude that considering only the fundamental TE$_{11}$ mode of circular waveguides offers a simple and \mbox{(semi-)} quantitative description of the phase data of small cylindrical apertures.

This is corroborated by data for $D$\,=\,0.2\,mm for different aperture thicknesses $d_{\rm ap}$ from 0.5\,mm to 1.5\,mm, see Fig.\ \ref{fig:refs_02}. 
Our data show that the cut-off frequency is given by $D$ but independent of $d_{\rm ap}$ (within the studied range of $d_{\rm ap}$). 
For these tiny apertures, produced with a mechanical drill, a few percent difference of the actual value of $D$ from the nominal value yields a small variation of $\omega_c$ among the three studied apertures. 
At the same time, the phase and thus the terahertz path length difference $L^{\rm ap}_D$ strongly depend on $d_{\rm ap}$, in agreement with Eq.\ (\ref{eq:DL}). At $\omega_c$, $|L^{\rm ap}_D|$ roughly reaches the aperture thickness $d_{\rm ap}$ in all three cases. Remarkably, the intensity is somewhat enhanced below $\omega_c$ for $d_{\rm ap}$\,=\,0.5\,mm as compared to the thicker apertures. 
This suggests that the coupling efficiency of propagating waves in front of and behind the small aperture is enhanced for a thinner aperture, as previously observed in the visible frequency range \cite{Degiron04} as well as in the terahertz range \cite{Zhang13}.

\begin{figure}[t!]
	\centering
	\includegraphics[width=\columnwidth]{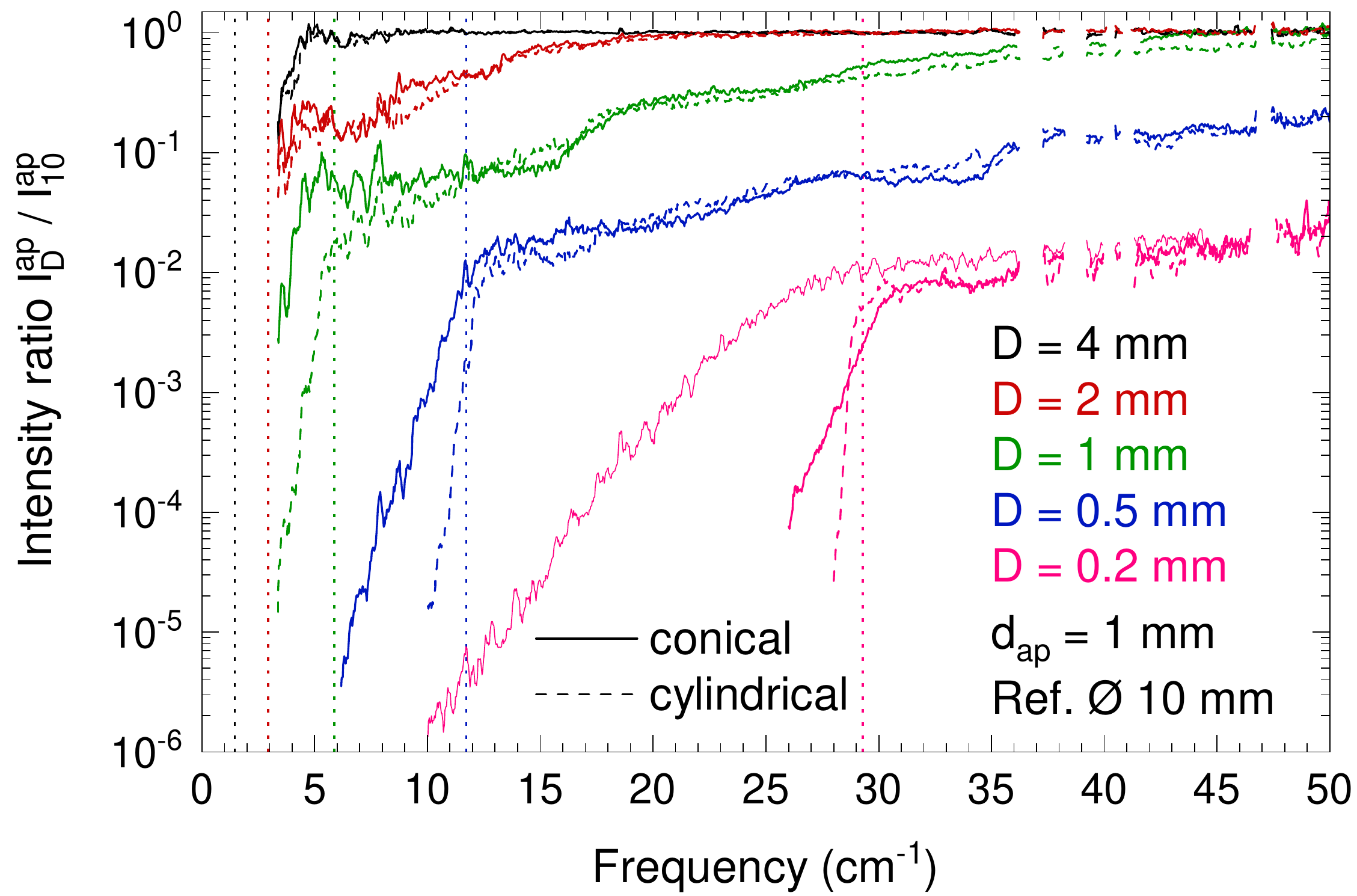}
	\includegraphics[width=\columnwidth]{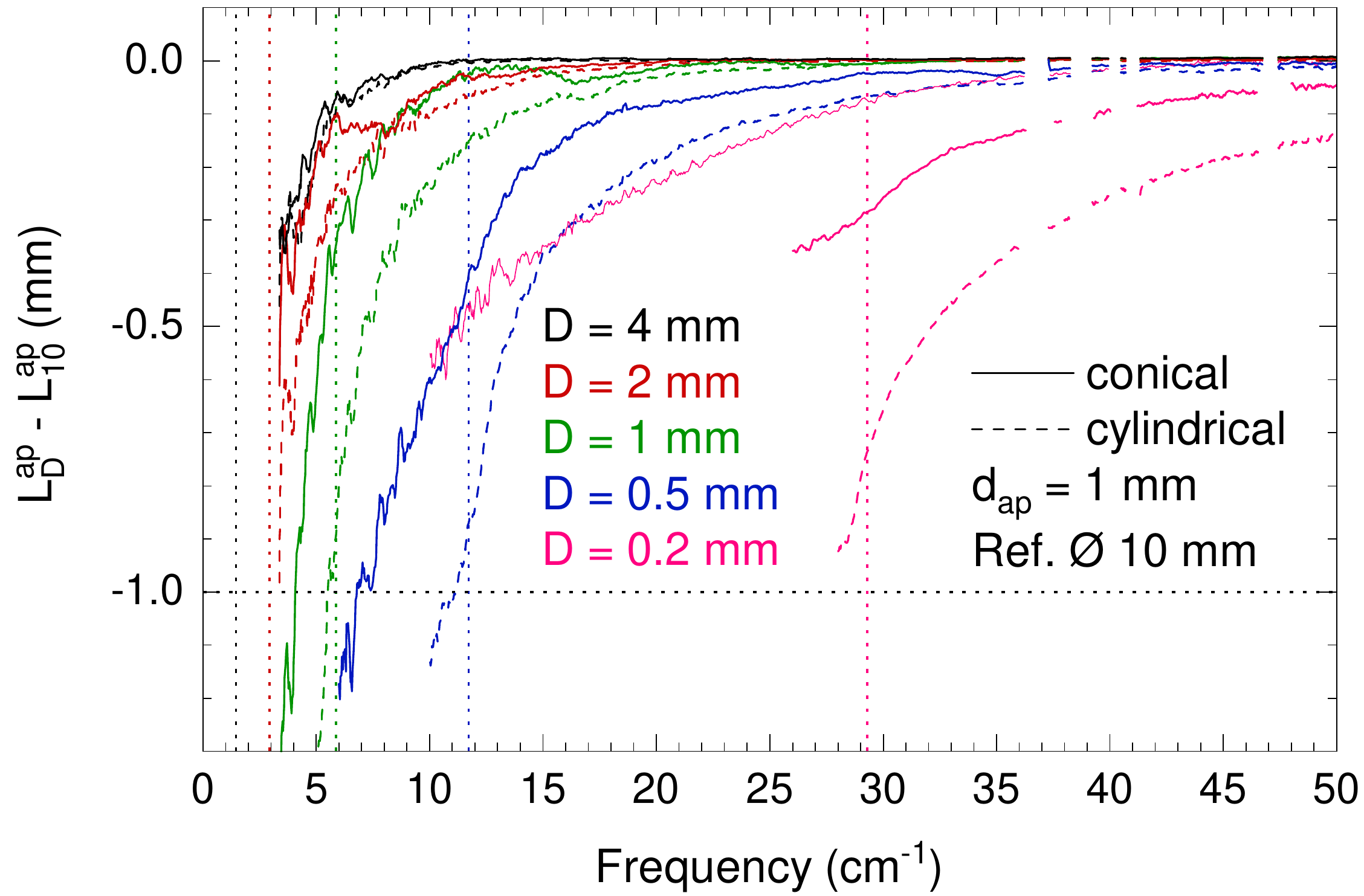}
	\caption{Intensity ratio $I^{\rm ap}_D/I^{\rm ap}_{10}$ (top) and terahertz path length difference $L^{\rm ap}_D-L^{\rm ap}_{10}$ (bottom) for apertures partially widened to a conical shape with different minimal diameters $D$ and thickness $d_{\rm ap}$\,=\,1\,mm, normalized to a reference with $D$\,=\,10\,mm. 
	For comparison, data of cylindrical bores are shown by dashed lines (cf.\ Fig.\ \ref{fig:refs_Trans_DeltaL}). For $D$\,=\,0.2\,mm, data for two partially conical apertures are depicted, featuring a cylindrical part with $d_{\rm cyl}$\,=\,$(0.11\pm 0.01)$\,mm (thin solid line) and 0.3\,mm, respectively.
	In both panels, vertical dotted lines denote the circular	
	waveguide cut-off frequency $\omega_c$ for the curves of the same color.  	
	In the bottom panel, the horizontal dotted line denotes $-d_{\rm ap}$. 
	Above 35\,cm$^{-1}$, the narrow features due to water vapor were cut out from the data. 
}
\label{fig:refs_cyl_vs_con}
\end{figure}

\subsection{Conical bore}
\label{sec:conicalbore}

A similar enhancement of the transmitted intensity below the cut-off frequency $\omega_c$ of a cylindrical waveguide is observed if the bore is partially widened to a cone on one face of the aperture, see top panel of Fig.\ \ref{fig:refs_cyl_vs_con}. For instance for $D$\,=\,0.5\,mm, the drop of the 
intensity below about 12\,cm$^{-1}$ is still pronounced but less steep than for the cylindrical aperture. 
Above $\omega_c$, the intensity is similar for conical and cylindrical bores. 
A conical bore can effectively be viewed as a cylindrical bore with varying diameter, equivalent to a cut-off frequency that varies along the path, $\omega_c^{\rm con}(z) \leq \omega_c$. As a consequence, a mode with $\omega < \omega_c$ may propagate some distance within the cone and then develop into an evanescent mode for the remaining portion of the aperture. 
This effectively reduces the thickness of the aperture and enhances the coupling efficiency to propagating modes behind the aperture. 
This picture is supported by comparing two partially conical apertures with $D$\,=\,0.2\,mm 
that differ in the length $d_{\rm cyl}$ of the remaining cylindrical part, see Fig.\ \ref{fig:refs_cyl_vs_con}. 
For $d_{\rm cyl}$\,=$(0.1\pm 0.01)$\,mm, the intensity below $\omega_c$ is considerably enhanced in comparison to data for $d_{\rm cyl} \approx 0.3$\,mm.

While the effect of a conical bore on $I^{\rm ap}_D/I^{\rm ap}_{10}$ is rather restricted to frequencies below $\omega_c$, 
the phase data of conical apertures differ over a broad frequency range from the results of cylindrical apertures, 
see Fig.\ \ref{fig:refs_cyl_vs_con}. In particular, we observe a clear enhancement of $L^{\rm ap}_D$ for conical apertures, both above and below $\omega_c$. In other words, the terahertz path length difference $L^{\rm ap}_D-L^{\rm ap}_{10}$ is closer to zero for a conical bore, i.e., the phase is less affected. 
As described above for the intensity, the conical shape effectively reduces the thickness of the aperture as well as the average phase velocity, which according to Eq.\ (\ref{eq:DL}) explains the observed enhancement of $L^{\rm ap}_D$.
Again, this point of view is supported by the data for $D$\,=\,0.2\,mm. 
The value of $L^{\rm ap}_{0.2}-L^{\rm ap}_{10}$ at $\omega_c$ roughly corresponds to the length $d_ {\rm cyl}$ of the cylindrical part of the aperture. The conical widening and the concomitant reduction of $d_{\rm cyl}$ shift $L^{\rm ap}_D-L^{\rm ap}_{10}$ closer to zero over the full measured frequency range, similar to the dependence on $d_{\rm ap}$ of cylindrical apertures depicted in the bottom panel of Fig.\ \ref{fig:refs_02}.

Overall, a conical bore extends the accessible range to lower frequencies. 
For, e.g., $D$\,=\,0.2\,mm and $d_{\rm ap} = 1$\,mm, the lower limit is shifted from about 30\,cm$^{-1}$ to about 10\,cm$^{-1}$, which is much smaller than $1/D$. The intensity, however, is strongly suppressed in this range. 
In practice, this shift of the lower limit heavily relies on the large dynamic range of our setup at low frequencies \cite{Roggenbuck10}.

\section{Lactose on small apertures}
\label{sec:lactose_small}

For transmission measurements of a given sample, the empty aperture is used as a reference. 
In the ideal case, the reference data cancels all effects of the aperture. 
This oversimplified picture assumes that the sample only affects amplitude and phase of the transmitted beam, 
with no effect on the direction of different rays. This works reasonably well for large apertures (one shortcoming 
is the neglect of the Gouy phase shift, see Appendix), but it fails when diffraction effects become important at low frequency. Further problems arise in practice if a small aperture suppresses the transmitted intensity too strongly.

\begin{figure}[t]
	\centering
	\includegraphics[width=\columnwidth]{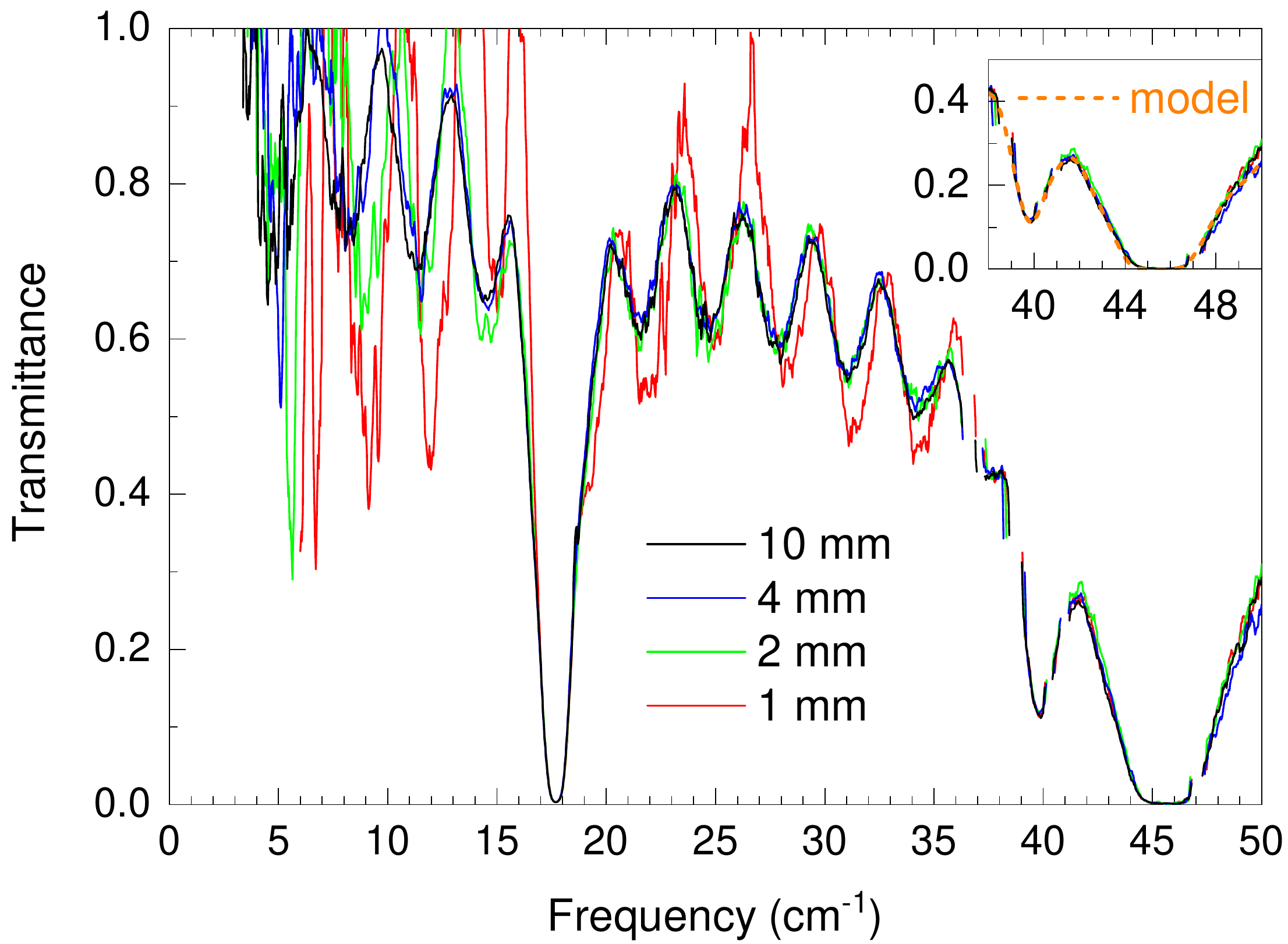}
    \includegraphics[width=\columnwidth]{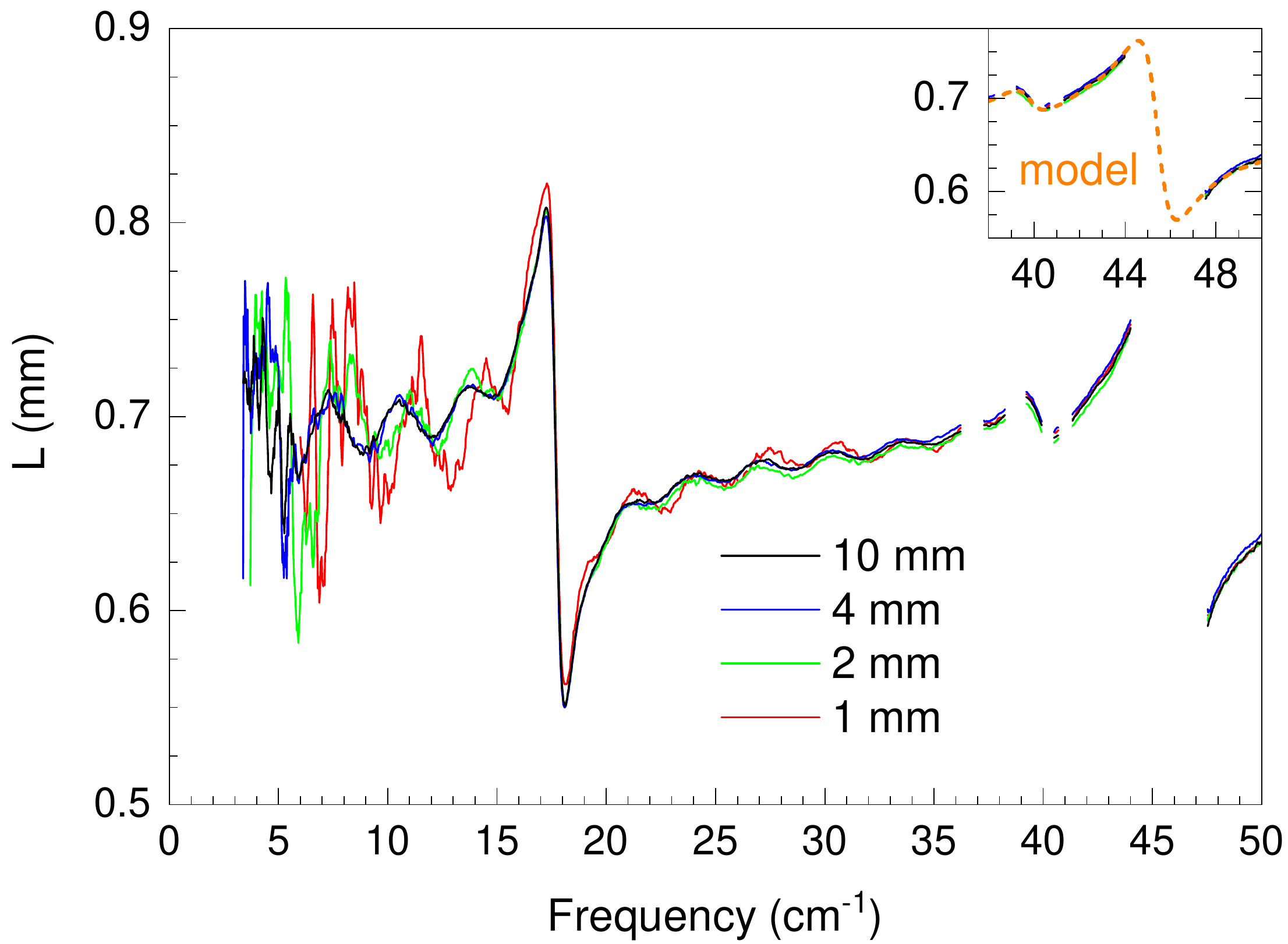}
	\caption{Transmittance $T(\omega)$ (top) and terahertz path length difference $L(\omega)$ (bottom) of a pressed pellet of $\alpha$-lactose monohydrate, measured with different conical apertures with minimum diameter $D \geq 1$\,mm as given in the figure. Data for $D$\,=\,10\,mm are taken from Fig.\ \ref{fig:Transmission_DeltaL_fit}. Insets additionally show the Drude-Lorentz fit as depicted in Fig.\ \ref{fig:Transmission_DeltaL_fit}.
	}
	\label{fig:DeltaL1mm}
\end{figure}

\begin{figure}[t]
	\centering
	\includegraphics[width=\columnwidth]{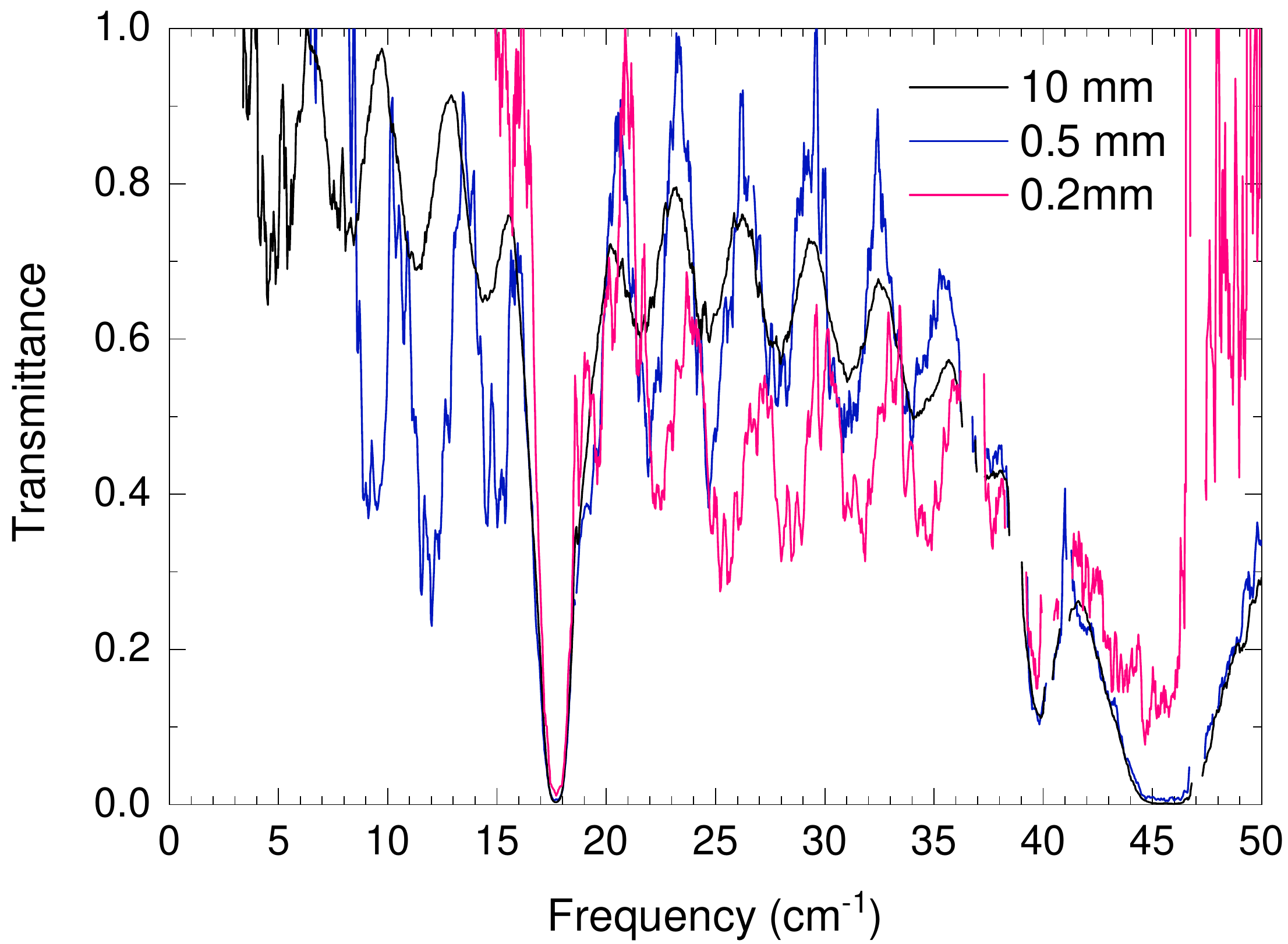}
	\includegraphics[width=\columnwidth]{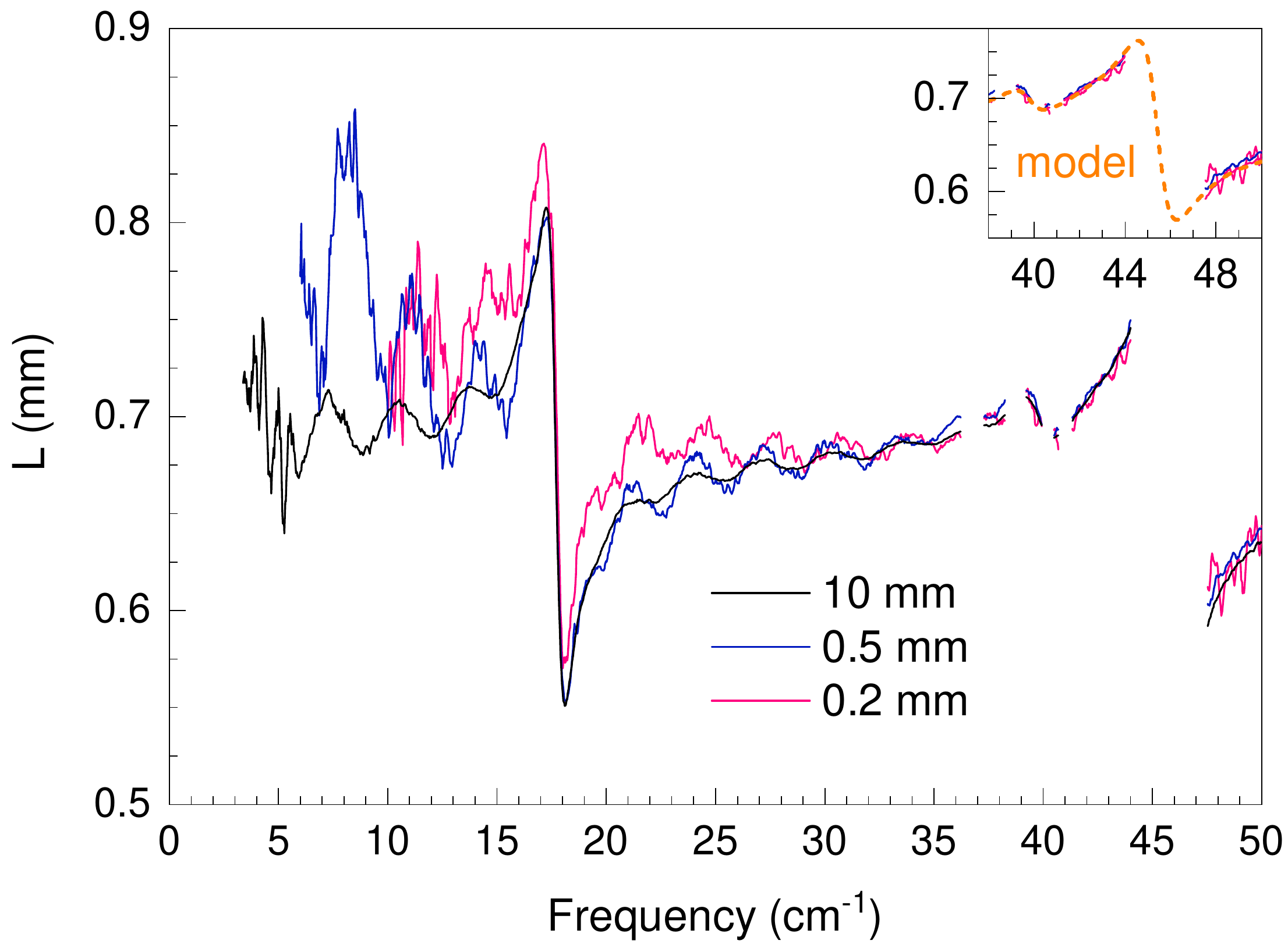}
	\caption{Transmittance $T(\omega)$ (top) and terahertz path length difference $L(\omega)$ (bottom) of a pressed pellet of $\alpha$-lactose monohydrate, measured with small partially conical apertures with overall thickness $d_{\rm ap}$\,=\,$1$\,mm. 
	The length of the cylindrical part equals $d_{\rm cyl}$\,=\,$(0.11\pm 0.01)$\,mm 
	for $D$\,=\,0.2\,mm and $d_{\rm cyl}$\,=\,$(0.21\pm 0.03)$\,mm for $D$\,=\,0.5\,mm, respectively.
	For comparison, data for $D$\,=\,10\,mm are taken from Fig.\ \ref{fig:Transmission_DeltaL_fit}. The inset additionally shows the Drude-Lorentz fit from Fig.\ \ref{fig:Transmission_DeltaL_fit}.
	}
	\label{fig:DeltaL0.2mm}
\end{figure}

Experimental data for a pressed pellet of $\alpha$-lactose monohydrate measured with conical apertures of minimum diameter $D$ down to 1\,mm are shown in Fig.\ \ref{fig:DeltaL1mm}. 
For clarity, the more noisy data of still smaller apertures are depicted in Fig.\ \ref{fig:DeltaL0.2mm}. 
The same sample pellet has been attached to the different apertures. For each data set, 
the bare aperture is employed for the reference measurement, 
and each curve represents the average of 20 measurements. 
Figure \ref{fig:DeltaL1mm} shows that the data for $D$\,=\,2\,mm agree very well with those for $D$\,=\,10\,mm  
down to about 10\,cm$^{-1}$, both for $T(\omega)$ and $L(\omega)$. 
As expected, deviations from the result for large $D$ increase with decreasing frequency and decreasing $D$.
For $D$\,=\,1\,mm, $T(\omega)$ shows considerable deviations from the $D$\,=\,10\,mm data below about 30\,cm$^{-1}$ and exceeds 1, i.e.\ becomes apparently unphysical, at 16\,cm$^{-1}$ and at further frequencies below. 
This demonstrates the limits for a simple quantitative evaluation of $T(\omega)$ for small samples 
at low frequencies. For $D$\,=\,0.5\,mm and 0.2\,mm, $T(\omega)$  exceeds 1 at about 8\,cm$^{-1}$ 
and 15\,cm$^{-1}$, respectively, and the observed result strongly deviates from the 10\,mm data over 
most of the measured frequency range, see Fig.\ \ref{fig:DeltaL0.2mm}.

\begin{figure}[t]
	\centering
	\includegraphics[width=\columnwidth]{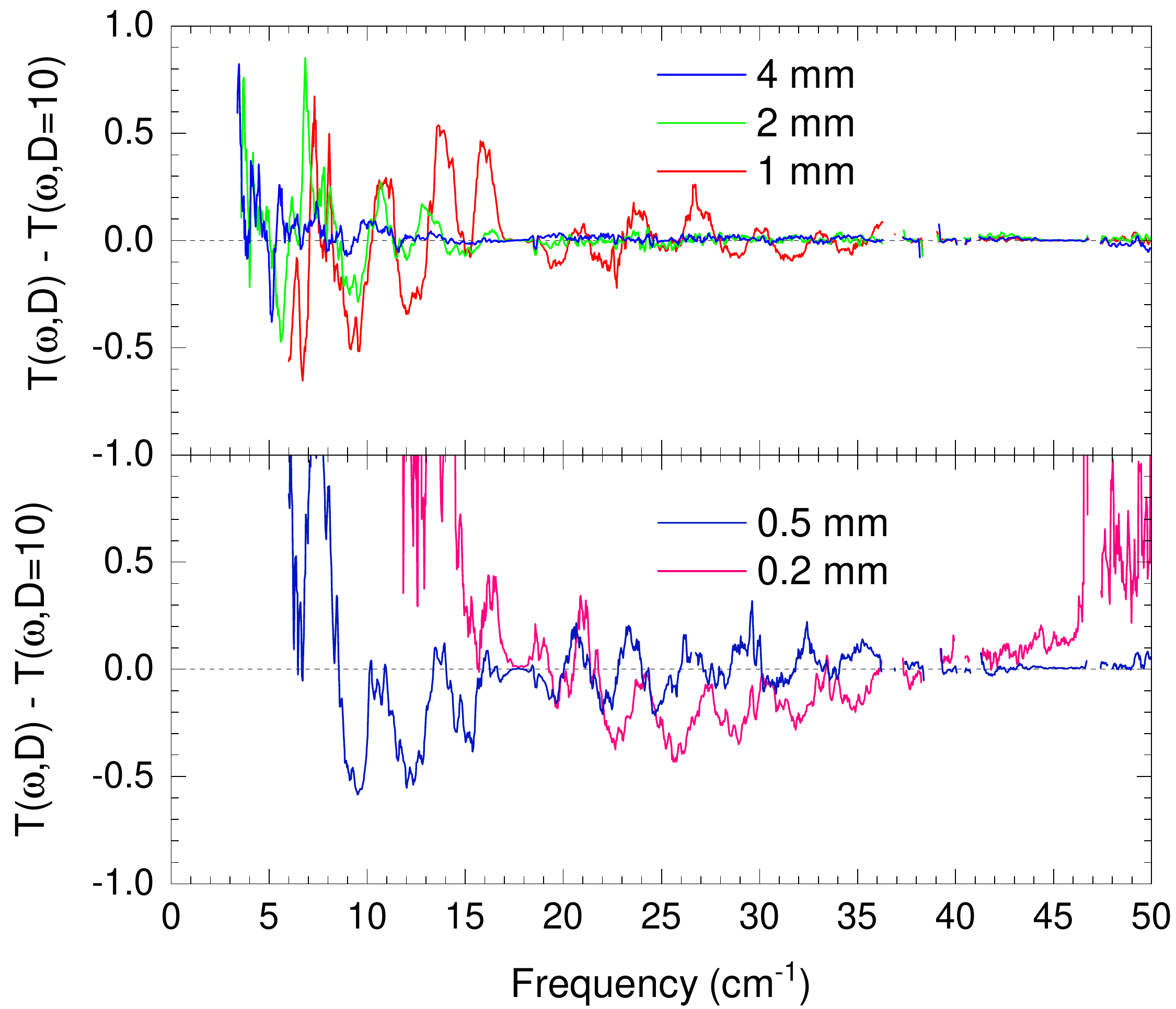}
	\includegraphics[width=\columnwidth]{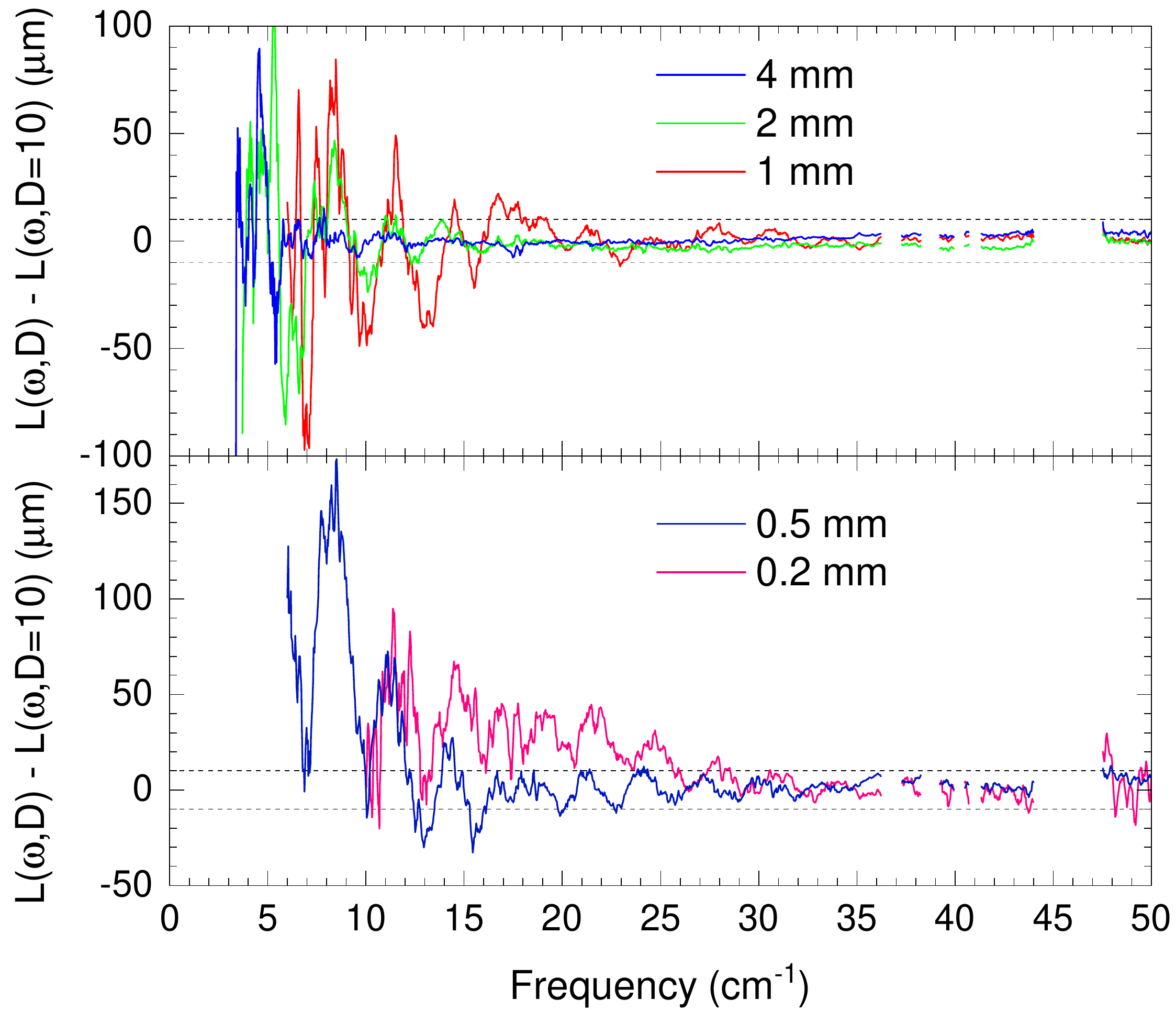}
	\caption{Differences $T(\omega,D)-T(\omega,D=10)$ (top panels) and $L(\omega,D)-L(\omega,D=10)$ (bottom) between data measured on $\alpha$-lactose monohydrate using an aperture diameter $D$ and the result for $D$\,=\,10\,mm, cf.\ Figs.\ \ref{fig:DeltaL1mm} and \ref{fig:DeltaL0.2mm}. 
    Dashed lines in the bottom panels indicate $\pm 10$\,$\mu$m. 
	}
	\label{fig:Ldiff}
\end{figure}

\begin{figure}[t]
	\centering
	\includegraphics[width=\columnwidth]{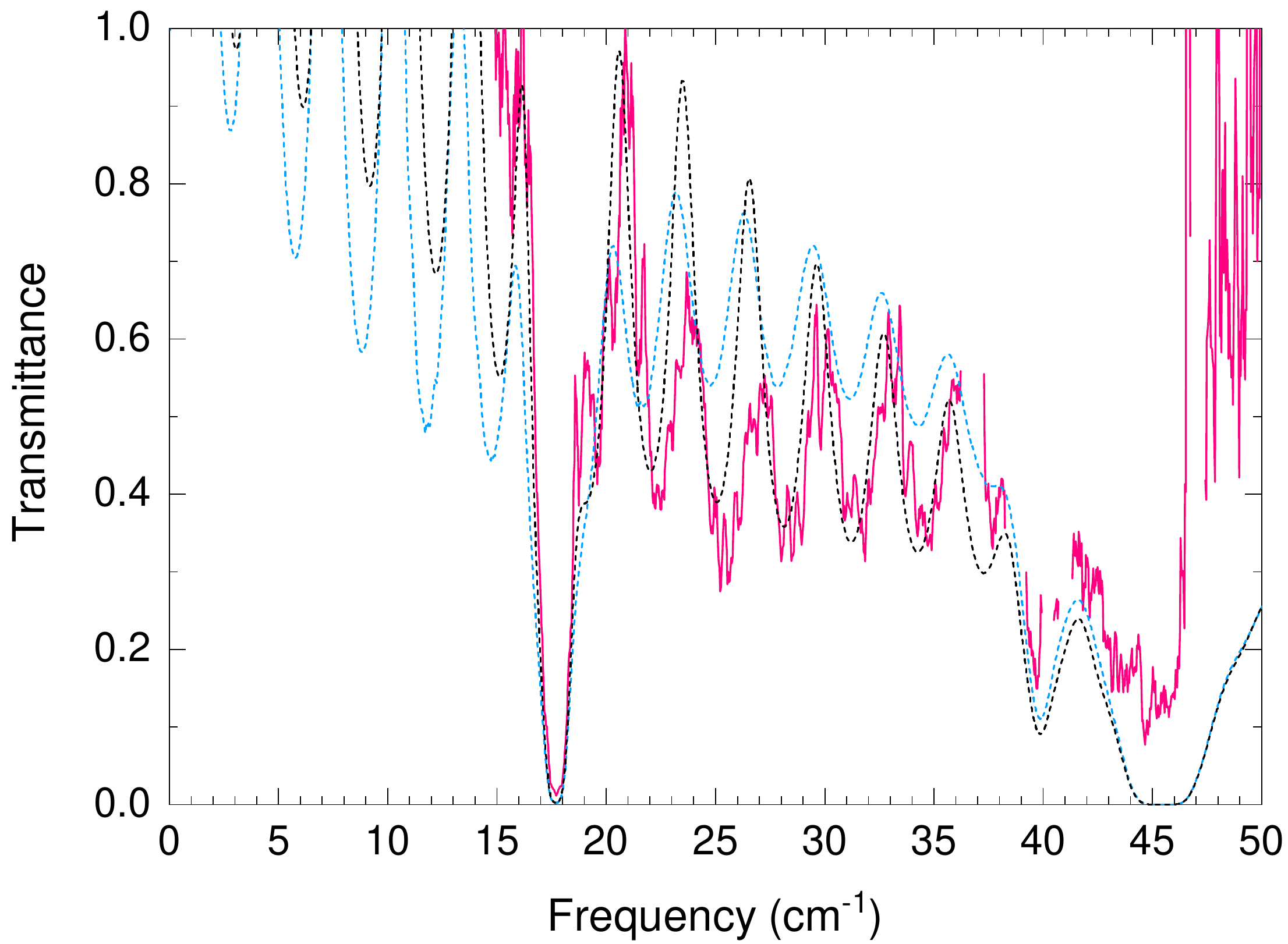}
	\includegraphics[width=\columnwidth]{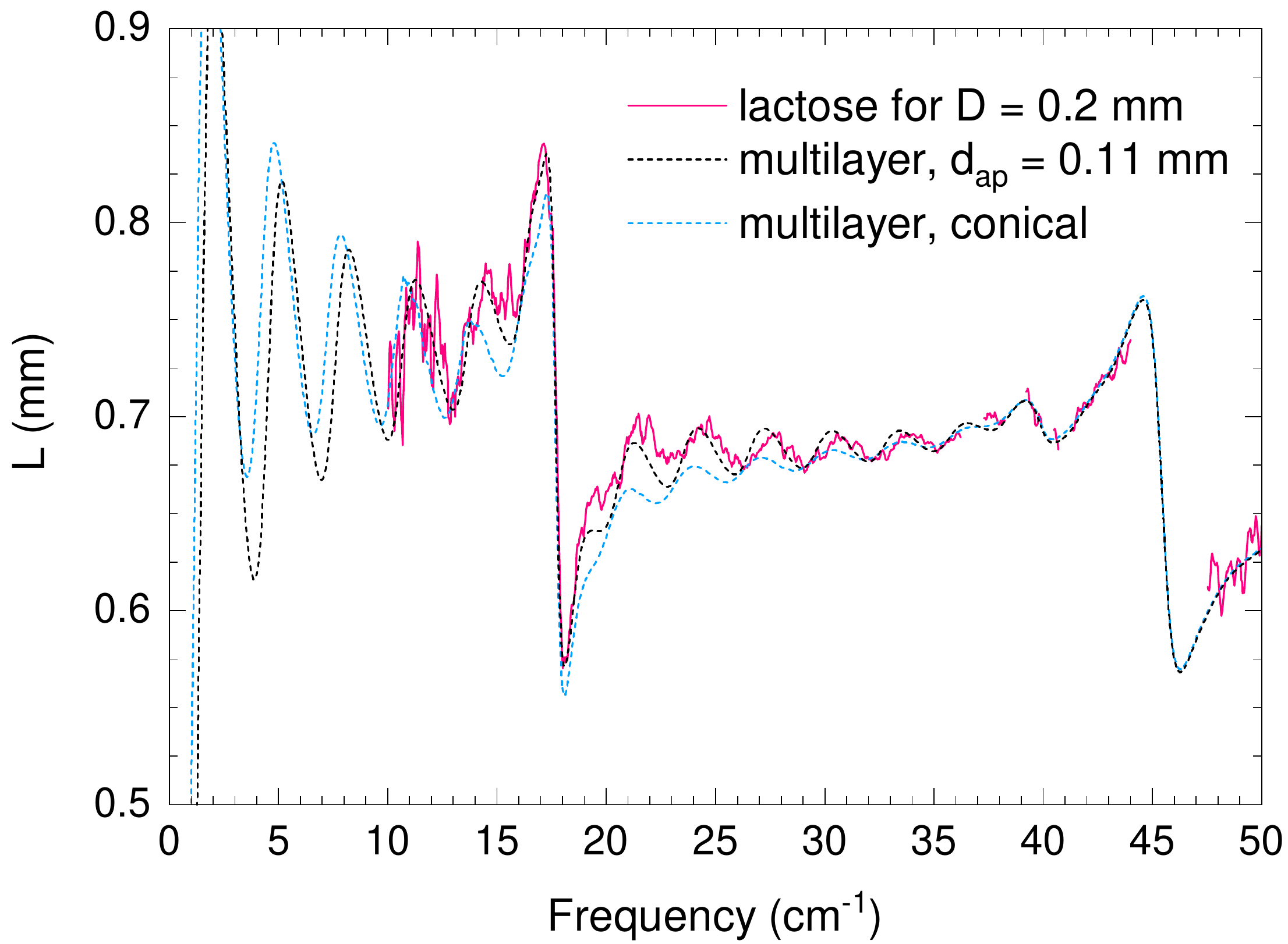}
	\caption{Comparison of the measured data for $D$\,=\,0.2\,mm (pink solid line) with 
		multilayer simulations (dashed). Black: stack of $\alpha$-lactose monohydrate and a circular 
		waveguide with $D$\,=\,0.2\,mm and $d_{\rm ap}$\,=\,0.11\,mm. 
		The blue line instead employs a conical waveguide with the measured properties, 
		cf.\ Fig.\ \ref{fig:refs_cyl_vs_con}.
		For $\alpha$-lactose monohydrate, we use the dielectric function determined from the 10\,mm data. 
	}
	\label{fig:multi}
\end{figure}

Also the terahertz path length difference $L(\omega)$ measured for small $D$ differs from the result for large $D$, see bottom panels of Figs.\ \ref{fig:DeltaL1mm} and \ref{fig:DeltaL0.2mm}. However, the analysis of $L(\omega)$ offers some advantages compared to $T(\omega)$. 
One example is the absorption mode peaking at 45\,cm$^{-1}$, which for $D$\,=\,0.2\,mm is very well captured in $L(\omega)$ in contrast to the case of $T(\omega)$. 
For all $D$ studied here, we cannot determine $L(\omega)$ in the narrow frequency range of anomalous dispersion above about 45\,cm$^{-1}$ where $L(\omega)$ decreases with increasing $\omega$, see Fig.\ \ref{fig:Transmission_DeltaL_fit} and insets of Figs.\ \ref{fig:DeltaL1mm} and \ref{fig:DeltaL0.2mm}.
This is due to the strong suppression of $T(\omega)$ for a sample thickness of 0.85\,mm. 
However, the observed behavior of $L(\omega)$ in the range of normal dispersion is sufficient to determine the properties of this absorption feature within a Drude-Lorentz oscillator model.
Furthermore, reasonable but somewhat noisy data of $L(\omega)$ can be measured down to about  6\,cm$^{-1}$ for both $D$\,=\,1\,mm and 0.5\,mm and down to about 10\,cm$^{-1}$ for $D$\,=\,0.2\,mm. 
In all three cases, this is lower than the cut-off in $T(\omega)$ which is given by $T>1$. 
For small $D$, Fig.\ \ref{fig:DeltaL0.2mm} indicates that the deviations from the 10\,mm data are less pronounced and more systematic in $L(\omega)$ than in $T(\omega)$. 
In the following, we support this statement by the results provided in Figs.\ \ref{fig:Ldiff} and \ref{fig:multi}.

The bottom panels of Fig.\ \ref{fig:Ldiff} explore the systematic behavior of $L(\omega)$ by highlighting the difference $L(\omega,D)-L(\omega,D=10)$ between $L(\omega)$ for a given $D$ and the result for $D$\,=\,10\,mm.
For the large aperture with $D$\,=\,4\,mm, the difference stays below 10\,$\mu$m above about 7\,cm$^{-1}$, see dashed lines in Fig.\ \ref{fig:Ldiff}. For a discussion of the sources of experimental errors in $L(\omega)$ of the order of 10\,$\mu$m, e.g.\ a thermal drift of the setup, we refer to Refs.\ \cite{Langenbach14,Komu15}. 
For $D$\,=\,2\,mm and 1\,mm, the error stays below 10\,$\mu$m above about 10\,cm$^{-1}$ and 20\,cm$^{-1}$, 
respectively. In both cases, the larger deviations are mainly caused by an enhanced modulation amplitude of the low-frequency Fabry-Perot interference fringes and a frequency shift of these low-frequency fringes, see Fig.\ \ref{fig:DeltaL1mm}. If one averages over the fringes, the difference $L(\omega,D)-L(\omega,D=10)$ stays remarkably small for both $D$\,=\,2\,mm and 1\,mm. The data for $D < 1$\,mm reveal an additional effect. On average, the experimental result for $L(\omega)$ for $D < 1$\,mm is too large, and the deviation increases with decreasing frequency, 
see bottom panel of Fig.\ \ref{fig:Ldiff}.

The corresponding difference $T(\omega,D)-T(\omega,D=10)$ is depicted in the top panels of Fig.\ \ref{fig:Ldiff}. As far as the Fabry-Perot fringes are concerned, the transmittance shows a similar frequency shift and enhanced modulation amplitude as observed in $L(\omega)$. However, $T(\omega)$ additionally exhibits pronounced changes of the absolute value. For $D$\,=\,0.2\,mm, $T(\omega)$ is suppressed in the range 20--35\,cm$^{-1}$ but enhanced below 20\,cm$^{-1}$. In contrast, $T(\omega)$ is strongly suppressed between about 8\,cm$^{-1}$ and 20\,cm$^{-1}$ for $D$\,=\,0.5\,mm. 
In the standard case of measurements of a small single crystal with a given size and unknown properties, a quantitative analysis of $T(\omega)$ at low frequencies provides an enormous challenge.

Diffraction strongly affects the measured terahertz amplitude, but the phase information of the detected photons still can be analyzed in a meaningful way. We tentatively attribute the more systematic $D$ dependence of $L(\omega)$ to the reduced coupling of light into a small waveguide-like bore. In particular, this coupling is affected by the presence of the sample. In general, an interference maximum corresponds to constructive interference of rays after multiple reflections within the sample. For large $D$, it occurs for $m\lambda$\,=\,$2nd$ with integer $m$. 
However, for a sample glued onto an aperture with a small bore, the coupling of electromagnetic waves from within the sample into the small bore is suppressed in comparison to a large aperture. Accordingly, the amplitude reflected back into the sample is enhanced, giving rise to an enhanced modulation amplitude of the fringes. Moreover, the reflection on the small bore adds a phase shift, which may explain the observed shift of the interference fringes.  
Furthermore, comparing Figs.\ \ref{fig:refs_cyl_vs_con} and \ref{fig:Ldiff}, the frequency dependence of 
the increase of $L(\omega)$ for small $D$ can be understood as due to a sample-induced reduction of 
the \textit{negative} contribution of the small apertures.

In order to test the interpretation that the behavior of $L(\omega)$ for small $D$ can be attributed to the interface between the sample and the waveguide-like aperture, we compare the data for $D$\,=\,0.2\,mm with multilayer simulations.
We calculate $T(\omega)$ and $L(\omega)$ for the 4-layer stack vacuum -- aperture -- sample -- vacuum, see Appendix. 
For this test, we use the known properties of the sample, i.e., we use the dielectric function determined from the 10\,mm data. 
A simulation that describes the aperture as a circular waveguide with $D$\,=\,0.2\,mm and $d_{\rm ap}$\,=\,0.11\,mm, as measured for the cylindrical part, yields a reasonable description of the data of $L(\omega)$, see Fig.\ \ref{fig:multi}.
In particular, the simulation captures the enhancement of $L(\omega)$ 
below about 30\,cm$^{-1}$, the frequency shift of the fringes, and their increased amplitude. 
Remarkably, this is not the case for a simulation employing the measured properties of the conical 
aperture, see blue line in Fig.\ \ref{fig:multi}. These measured properties, depicted in 
Fig.\ \ref{fig:refs_cyl_vs_con}, effectively represent the behavior averaged over the conical and 
the cylindrical part of the aperture. Our simulations hence demonstrate that the properties of the 
interface between the sample and the aperture are essential for a quantitative description of the data 
for small $D$. For future work, this suggests that the multilayer approach is suitable to determine or at least estimate $\varepsilon(\omega)$ of a given small sample with unknown properties, e.g., by fitting the data of $L(\omega)$ while using an oscillator model for the sample layer.

For the transmittance, the simulation treating the aperture as a circular waveguide reproduces the suppression of  $T(\omega)$ around 25\,cm$^{-1}$ only in the minima of the fringes, it fails to properly describe the reduced average value of $T(\omega)$ in this range. 
Both simulations show that $T(\omega)$ exceeds 1 at low frequencies. The simulations use an empty 
aperture as reference, and the transmittance of the aperture is strongly suppressed at low frequencies 
where propagating modes are not supported. A transmittance larger than 1 is achieved if the presence 
of the sample enhances the coupling of electromagnetic modes in front of and behind the aperture.

\section{Conclusions}
\label{sec:conclusions}

We studied the terahertz properties of $\alpha$-lactose monohydrate using small apertures of different diameter $D$. For $D \geq 4$\,mm, the properties can be determined very accurately over the full range of our setup, i.e., from 3\,cm$^{-1}$ to 50\,cm$^{-1}$. For smaller $D$, the data quality gradually deteriorates towards low frequencies. We find enhanced deviations in both the transmittance $T(\omega)$ and the terahertz path length difference $L(\omega)$, the latter being derived from the phase data. Our results demonstrate that for small apertures with $D < 1$\,mm, the analysis of the phase offers advantages over the conventional study of the transmittance. In particular, the deviations in the phase data are more systematic and can be reasonably 
described by a simple multilayer model that treats the aperture as a circular waveguide. Such a multilayer 
analysis of the phase data hence allows one to determine the terahertz properties of small samples 
down to low frequencies.

Furthermore, we studied the cw terahertz properties of small cylindrical apertures. These show three distinct frequency ranges. At very low frequencies, below the cut-off frequency $\omega_c$ of the TE$_{11}$ mode, the apertures show evanescent modes but do not support propagating modes. 
Around $\omega_c$, this yields a reduction of the terahertz path length by $d_{\rm ap}$, the thickness of the aperture. 
At very high frequencies, apertures with $D \geq 2$\,mm do not affect the amplitude and the phase of the electromagnetic waves. This range can be described by free-beam optics. 
In the intermediate frequency range, the amplitude may strongly be reduced for small apertures, 
while the phase or, equivalently, the terahertz path length shows strong dispersion and an enhanced phase velocity which agrees very well with simple expectations for a single mode in a circular waveguide. 
Using conical apertures pushes the accessible frequency range to lower frequencies. 
Nevertheless a quantitative analysis mainly requires a detailed understanding of the cylindrical part 
of the aperture. 

Overall, the central parameter determining the lower frequency limit is the ratio $D/w_0$. A setup with reduced beam waist $w_0$ hence allows to study smaller samples. However, in the employed four-lens setup a smaller $w_0$ corresponds to thicker lenses and hence larger absorption, i.e., a smaller terahertz signal. Furthermore, the choice of the focal length $f$ may be restricted, e.g., by the dimensions of a cryostat for low-temperature measurements. Beyond the reduction of the beam waist $w_0$, the precise analysis of the phase data is a promising route to study still smaller samples.

\begin{acknowledgements}
We acknowledge fruitful discussions with M. Langenbach, A. Roggenbuck, and A. Deninger 
as well as funding from the Deutsche Forschungsgemeinschaft (DFG, German Research Foundation) via Project number 277146847 - CRC 1238 (project B02). 

\end{acknowledgements}

\section*{APPENDIX A: Gouy phase shift}

\begin{figure}[t]
	\centering
	\includegraphics[width=\columnwidth]{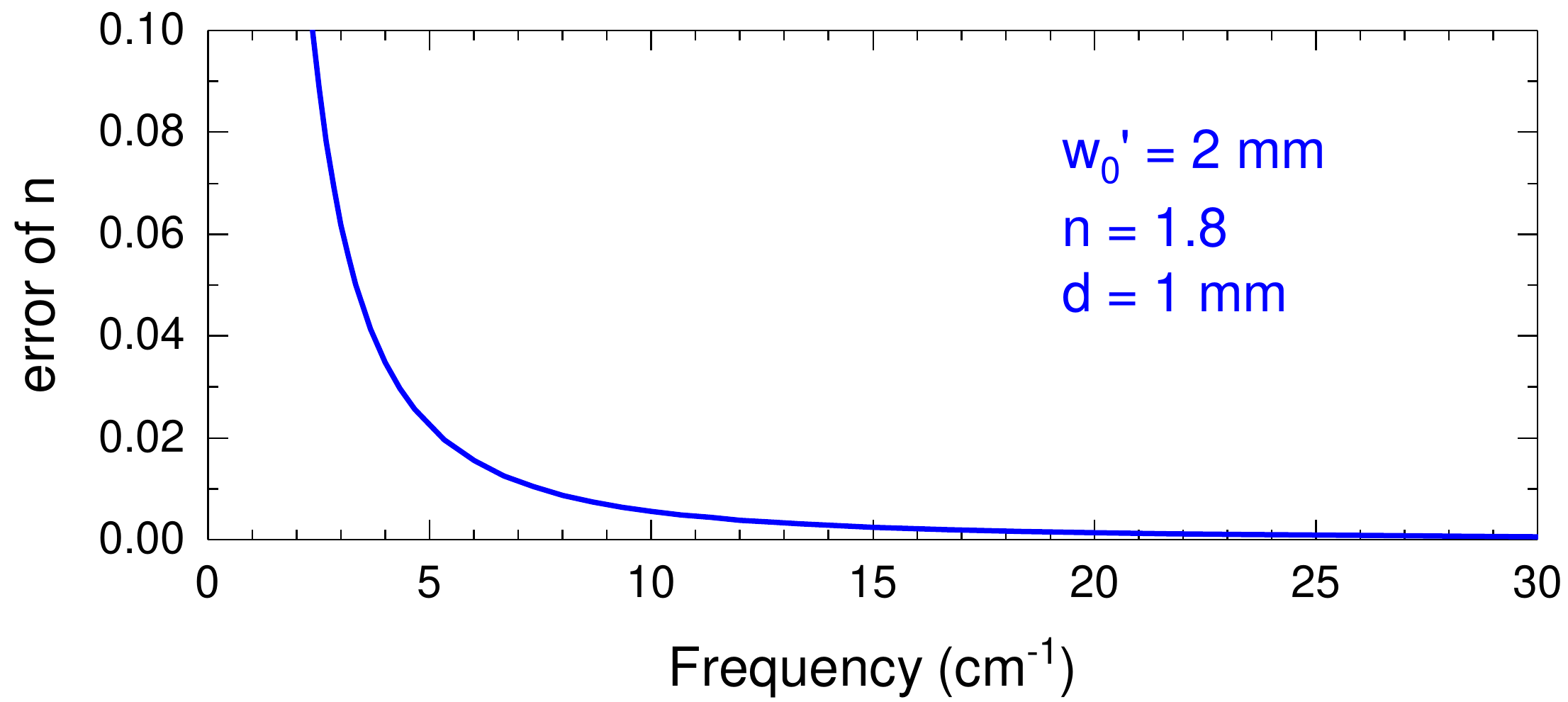}
	\caption{Error of $n$, i.e., the difference $n_{\rm eval}-n_{\rm true}$ (cf.\ Equ.\ \ref{eq_neval}), due to the Gouy phase shift for a 4-lens setup. The estimate uses the result of Ref.\ \cite{Liang15} with the given parameters.
	}
	\label{fig:error_n_Liang}
\end{figure}

The Gouy phase shift refers to a Gaussian beam passing through a beam waist in comparison to a plane wave. 
Its effect on the determination of the complex refractive index has been discussed 
quantitatively by Ku\v{z}el \textit{et al}.\ \cite{Kuzel10} for time-domain THz spectroscopy. 
The setup is first optimized for the reference measurement. Adding the sample shifts the focus which causes a corresponding shift of the focus at the detector position, giving rise to an error in the determined absorption and phase if they are analyzed in the standard way, neglecting the Gouy phase shift. 
A quantitative analysis for a 4-lens setup was reported by Liang \textit{et al}.\ \cite{Liang15}. 
They find that the difference between the true $n_{\rm true}$  and the actually obtained $n_{\rm eval}$ amounts to 
\begin{equation}
\label{eq_neval}
n_{\rm eval}-n_{\rm true} = \frac{c}{2\pi \nu d} \arctan\left[ 
\frac{(n_{\rm true}-1)d}{n_{\rm true}z_0'}\right]
\end{equation}
with
\begin{equation}
z_0'  = \frac{\pi {w_0'}^2 \nu}{c}
\end{equation}
where $w_0'$ denotes the radius for the power integration. 
For an estimate, we use $w_0'$\,=\,2\,mm, $n$\,=\,1.8 (for Lactose), and $d$\,=\,1\,mm. 
The estimated error is small, below 0.06 (0.01) above about 3\,cm$^{-1}$ (8\,cm$^{-1}$), 
see Fig.\ \ref{fig:error_n_Liang}. 
This agrees with the order of magnitude of the correction discussed experimentally in 
Ref.\ \cite{Liang15}. This error due to the Gouy phase shift is relevant for \textit{large} samples 
where one can achieve very precise results. For small samples, the challenge is to get a reasonable result at all. 
The Gouy phase shift is negligible in this case.

\section*{APPENDIX B: transmittance for small apertures}

For large apertures, the transmittance is evaluated assuming a free-standing slab of material with interfaces to vacuum on both sides. The analysis is based on the Fresnel coefficients for an interface between media 1 and 2 for normal incidence, 
\begin{eqnarray}
r_{12} & = & -r_{21} = \frac{N_1-N_2}{N_1+N_2} = \frac{Z_2-Z_1}{Z_1+Z_2}
\\
t_{12} & = & \frac{2 N_1}{N_1+N_2} = \frac{2 Z_2}{Z_1+Z_2}
\\
t_{21} & = & \frac{2 N_2}{N_1+N_2} = \frac{2 Z_1}{Z_1+Z_2}
\end{eqnarray}
where $N_i$ and $Z_i$ denote the complex refractive index and the wave impedance, respectively, 
with $N_i^2$\,=\,$\varepsilon_i$. 
For a free-standing slab with refractive index $N$ and thickness $d$, the ratio of the transmitted electric field to the incident one is given by 
\begin{equation}
\frac{E_t}{E_i} = \frac{4 N \exp(i q d)}{(N+1)^2 - (N-1)^2 \exp(i 2 q d)} \, .
\end{equation}

For a quantitative description of the data measured with small apertures, we consider a multilayer approach and calculate the transmitted electric field for the system vacuum - aperture - sample - vacuum (v-a-s-v). For such a 4-layer system, one finds \cite{Dressel}
\begin{equation}
\nonumber 
\frac{E_t}{E_i} = \frac{  t_{va} t_{as} t_{sv} e^{i(q_{\rm ap} d_{\rm ap} + q_{\rm s} d)} }
{1 - r_{sa} r_{sv} e^{i2 q_{\rm s} d} - r_{av} e^{i2 q_{\rm ap} d_{\rm ap}} (r_{as} + r_{sv} e^{i2 q_{\rm s} d}) }  \,\,\, ,
\end{equation}
where $d_{\rm ap}$ is the effective thickness of the aperture and $q_i$\,=\,$\frac{2\pi}{\lambda}N_i$. 
This approach requires an expression for the wave impedance $Z_{\rm ap}$ of the aperture. 
For a TE mode in a circular wave\-guide, the wave impedance 
is given by \cite{Gallot2000}
\begin{equation}
Z_{\rm w} = Z_0 \left(1-\left( \frac{\omega_c}{\omega}\right)^2 \right)^{-1/2} 
\end{equation}
with the vacuum impedance $Z_0$.

\end{document}